\documentclass[nojss]{jss}
\usepackage{amsmath,amsthm,float,orcidlink,thumbpdf,lmodern}
\graphicspath{{Figures/}}
\shortcites{Bezanson2018,scipy,stan,pyro,scibior2017denotational}

\usepackage[utf8]{inputenc}
\usepackage[T1]{fontenc}
\usepackage{jlcode}

\theoremstyle{definition}
\newtheorem{example}{Example}[section]

\author{Mathieu Besan\c{c}on~\orcidlink{0000-0002-6284-3033}\\Zuse Institute Berlin
   \And Theodore Papamarkou~\orcidlink{0000-0002-9689-543X}\\ The University of Manchester
   \And David Anthoff~\orcidlink{0000-0001-9319-2109}\\University of \\California at Berkeley
   \AND Alex Arslan\\ Beacon Biosignals, Inc.
   \And Simon Byrne~\orcidlink{0000-0001-8048-6810}\\ California Institute \\ of Technology
   \And Dahua Lin~\orcidlink{0000-0002-8865-7896}\\ Chinese University \\ of Hong Kong
   \And John Pearson~\orcidlink{0000-0002-9876-7837}\\ Duke University \\ Medical Center
   }
\Plainauthor{Mathieu Besan\c{c}on, David Anthoff, Alex Arslan, Simon Byrne, Dahua Lin, Theodore Papamarkou, John Pearson}

\title{\pkg{Distributions.jl}: Definition and Modeling of Probability Distributions in the JuliaStats Ecosystem}
\Plaintitle{Distributions.jl: Definition and Modeling of Probability Distributions in the JuliaStats Ecosystem}
\Shorttitle{\pkg{Distributions.jl}: Probability Distributions in the JuliaStats Ecosystem}

\Abstract{
  Random variables and their distributions are a central part in many areas of statistical methods.
  The \pkg{Distributions.jl} package provides \proglang{Julia} users and developers tools for working with probability
  distributions, leveraging \proglang{Julia} features for their intuitive and flexible manipulation, while remaining
  highly efficient through zero-cost abstractions.
}

\Plainkeywords{Julia, distributions, modeling, interface, mixture, KDE, sampling, probabilistic programming, inference}

\Volume{98}
\Issue{16}
\Month{July}
\Year{2021}
\Submitdate{2019-03-19}
\Acceptdate{2020-11-15}
\DOI{10.18637/jss.v098.i15}

\Address{
  Mathieu Besan\c{c}on\\
  Zuse Institute Berlin\\
  \emph{and} \\
  Department of Applied Mathematics \& Industrial Engineering\\
  Polytechnique Montr{\'e}al, Montr{\'e}al, QC, Canada\\
  \emph{and} \\
  {\'E}cole Centrale de Lille, Villeneuve d'Ascq, France\\
  \emph{and} \\
  INOCS, INRIA Lille \& CRISTAL, Villeneuve d'Ascq, France\\
  E-mail: \email{mathieu.besancon@polymtl.ca}\\

  Theodore Papamarkou\\
  Department of Mathematics\\
  The University of Manchester\\
  \emph{and} \\
  Department of Mathematics\\
  University of Tennessee\\
  \emph{and} \\
  Computational Sciences \& Engineering Division\\
  Oak Ridge National Laboratory\\
  E-mail: \email{theodore.papamarkou@manchester.ac.uk}\\

  David Anthoff\\
  Energy and Resources Group\\
  University of California at Berkeley\\
  E-mail: \email{anthoff@berkeley.edu}\\

  Alex Arslan \\
  Beacon Biosignals, Inc. \\
  E-mail: \email{alex.arslan@beacon.bio}\\

  Simon Byrne\\
  California Institute of Technology\\
  E-mail: \email{simonbyrne@caltech.edu}\\

  Dahua Lin\\
  The Chinese University of Hong Kong\\
  E-mail: \email{dhlin@ie.cuhk.edu.hk}\\

  John Pearson\\
  Department of Biostatistics and Bioinformatics\\
  Duke University Medical Center\\
  Durham, NC, United States of America\\
  E-mail: \email{john.pearson@duke.edu}
}

\begin{document}

\section{Introduction}\label{sec:intro}

The \pkg{Distributions.jl} package \citep{distributions} defines interfaces for
representing probability distributions and other mathematical objects describing
the generation of random samples. It is part of the JuliaStats open-source
organization hosting packages for core mathematical and statistical functions
(\pkg{StatsFuns.jl}, \pkg{PDMats.jl}), model fitting
(\pkg{GLM.jl} for generalized linear models, \pkg{HypothesisTests.jl}, \pkg{MixedModels.jl},
\pkg{KernelDensity.jl}, \pkg{Survival.jl}) and utilities (\pkg{Distances.jl}, \pkg{Rmath.jl}).

\pkg{Distributions.jl} defines generic and specific behavior required for
distributions and other ``sample-able'' objects. The package implements a large
number of distributions to be directly usable for probabilistic modeling,
estimation and simulation problems. It leverages idiomatic features of
\proglang{Julia} including multiple dispatch and the type system, both presented
in more detail in \cite{julia101} and \cite{Bezanson2018}.

In many applications, including but not limited to physical
simulation, mathematical optimization or data analysis,
computer programs need to generate random numbers from sample spaces
or more specifically from a probability distribution or to infer a
probability distribution given prior knowledge and observed samples.
\pkg{Distributions.jl} unifies these use cases under one set of modeling tools.

A core idea of the package is to build an equivalence between mathematical
objects and corresponding \proglang{Julia} types. A probability distribution
is therefore represented by a type with a given behavior provided through
the implementation of a set of methods. This equivalence makes the syntax
of programs fit the semantics of the mathematical objects.

\subsection*{Related software}

In the \code{scipy.stats}\footnote{\url{https://docs.scipy.org/doc/scipy/reference/stats.html}}
module of the \pkg{SciPy} project \citep{scipy}, distributions
are created and manipulated as objects in the
object-oriented programming sense.
Methods are defined for a distribution class,
then specialized for continuous and discrete distribution classes
inheriting from the generic distribution, from which inherit
classes representing actual distribution families.

Representations of probability distributions have also been
implemented in statically-typed functional programming languages,
in \proglang{Haskell} in the \pkg{Probabilistic Functional Programming}
package \citep{haskell}, and in \proglang{OCaml} \citep{funcprobprog},
supporting only discrete distributions as an association
between collection elements and probabilities.
The work in \cite{scibior2015practical} presents a generic
monad-based representation of probability distributions allowing
for both continuous and discrete distributions.

The \pkg{stats} package, which is distributed as part of the
\proglang{R} language, includes functions related to probability distributions
which use a prefix naming convention: \code{r\emph{dist}} for random sampling, \code{d\emph{dist}} for computing the probability density, \code{p\emph{dist}} for computing the cumulative distribution function, and \code{q\emph{dist}} for computing the quantiles.
The \pkg{distr} package \citep{distr} also allows \proglang{R} users
to define their own distribution as a class of the S4 object-oriented system,
with four functions \code{r, d, p, q} stored by the object.
Only one of the four functions has to be provided by the user when creating
a distribution object, the other functions are computed in a suitable way.
This approach increases flexibility but implies a runtime cost
depending on which function has been provided to define a given distribution object.
For instance, when only the random generation function is provided,
the \code{RtoDPQ} function empirically constructs an estimation for the others,
which requires drawing samples instead of directly evaluating an
analytical density.

In \proglang{C++}, the \pkg{boost} library \citep{boost}, the  \code{maths}
component includes common distributions and computations upon them.
As in \pkg{Distributions.jl}, probability distributions are represented by
types (classes), as opposed to functions. The underlying numeric types are
rendered generic by the use of templates.
The parameters of each distribution are accessed through exposed methods,
while common operations are defined as non-member functions, thus sharing a similar
syntax with single dispatch.\footnote{\url{https://www.boost.org/doc/libs/1_69_0/libs/math/doc/html/math_toolkit/stat_tut/overview/generic.html}}
Design and implementation proposals for
a multiple dispatch mechanism in \proglang{C++} have been investigated in
\cite{cppdispatch} and described as a library in \cite{LEGOC2015531},
which would allow for more sophisticated dispatch rules in the \pkg{Boost}
distribution interface.

This paper is organized as follows.
Section \ref{sec:types} presents the main types defined in the package
and their hierarchy. In Section \ref{sec:sampling},
the \code{Sampleable} type and the associated sampling interface are presented.
Section \ref{sec:distribution} presents the distribution interface,
which is the central part of the package.
Section \ref{sec:fitting} presents the available tools for fitting
and estimation of distributions from data using parametric and non-parametric techniques.
Section \ref{sec:mixtures} presents modeling tools and algorithms for
mixtures of distributions.
Section \ref{sec:applications} highlights two applications of
\pkg{Distributions.jl} in related packages for kernel density estimation
and the implementation of Probabilistic Programming Languages in pure \proglang{Julia}.
Section \ref{sec:conclusion} concludes
on the work presented and on future development of the ecosystem
for probability distributions in \proglang{Julia}.

\section{Type hierarchy}\label{sec:types}

The \proglang{Julia} language allows the definition of new types and their use
for specifying function arguments using the multiple dispatch mechanism \citep{ZappaNardelli18}.

Most common probability distributions can be broadly classified along two facets:
\begin{itemize}
    \item the dimensionality of the values 
    (e.g., univariate, multivariate, matrix variate)
    \item whether it has \emph{discrete} or \emph{continuous} support,
    corresponding to a density with respect to a counting measure or a Lebesgue measure
\end{itemize}
In the \proglang{Julia} type system semantics, these properties can be captured by adding type
parameters characterizing the random variable to the distribution type which
represents it. Parametric typing makes these pieces of information on the sample
space available to the \proglang{Julia} compiler, allowing dispatch to be performed at
compile-time, making the operation a zero-cost abstraction.

\code{Distribution} is an abstract type that takes two parameters: a \code{VariateForm} type which describes the dimensionality, and \code{ValueSupport} type which describes the discreteness or continuity of the support.
These ``property types'' have singleton subtypes which enumerate these properties:
\begin{lstlisting}[language=Julia]
julia> abstract type VariateForm end
julia> struct Univariate    <: VariateForm end
julia> struct Multivariate  <: VariateForm end
julia> struct MatrixVariate <: VariateForm end
julia> abstract type ValueSupport end
julia> struct Discrete   <: ValueSupport end
julia> struct Continuous <: ValueSupport end
\end{lstlisting}
Various type aliases are then defined for user convenience:
\begin{lstlisting}[language=Julia]
julia> DiscreteUnivariateDistribution     = Distribution{Univariate, Discrete}
julia> ContinuousUnivariateDistribution   = Distribution{Univariate, Continuous}
julia> DiscreteMultivariateDistribution   = Distribution{Multivariate, Discrete}
julia> ContinuousMultivariateDistribution = Distribution{Multivariate, Continuous}
\end{lstlisting}
The \proglang{Julia} \code{<:} operator in the definition of a new
type specifies the direct supertype.
Specific distribution families are then implemented as
sub-types of \code{Distribution}:
typically these are defined as composite types ("\code{struct}")
with fields capturing the parameters of the distribution.
Further information on the type system can be found in Appendix \ref{sec:apptype}.
For example, the univariate uniform distribution on
$(a,b)$ is defined as:
\begin{lstlisting}[language=Julia]
julia> struct Uniform{T<:Real} <: ContinuousUnivariateDistribution
    a::T
    b::T
end
\end{lstlisting}
Note in this case the \code{Uniform} distribution is itself a parametric type,
this allows it to make use of different numeric types. By default these are
\code{Float64}, but they can also be \code{Float32}, \code{BigFloat},
\code{Rational}, the \code{Dual} type from \pkg{ForwardDiff.jl} in order to
support features like automatic differentiation \citep{revels2016forward}, or user defined number types.

Probabilities are assigned to subsets in a sample space,
probabilistic types are qualified based on this sample space
from a \code{VariateForm} corresponding to ranks of the samples
(scalar, vector, matrix, tensor)
and a \code{ValueSupport} corresponding to the set from
which each scalar element is restricted.

Other types of sample spaces can be defined
for different use cases by implementing new sub-types.
We provide two examples below, one for \code{ValueSupport}
and one for \code{VariateForm}.

There are different possibilities to represent stochastic processes
using the tools from \pkg{Distributions.jl}. One possibility is to
define them as a new \code{ValueSupport} type.
\begin{lstlisting}[language=Julia]
julia> using Distributions
julia> abstract type StochasticProcess <: ValueSupport end
julia> struct ContinuousStochasticProcess <: StochasticProcess end
julia> abstract type ContinuousStochasticSampler{F<:VariateForm}
<: Sampleable{F,ContinuousStochasticProcess}
end
\end{lstlisting}
More complete examples of representations of stochastic processes can be
found in \pkg{Bridge.jl} \citep{bridge}.
\pkg{Distributions.jl} can also be extended to support tensor-variate random variables in a similar fashion:
\begin{lstlisting}[language=Julia]
julia> using Distributions
julia> struct TensorVariate <: VariateForm end
julia> abstract type TensorSampleable{S<:ValueSupport}
<: Sampleable{TensorVariate,S}
end
\end{lstlisting}
This allows other developers to define their own models on
top of \pkg{Distributions.jl} without requiring the modification
of the package, while end-users benefit from the same interface
and conventions, regardless of whether one type was defined
in \pkg{Distributions.jl} or in an external package.

The types describing a probabilistic sampling process then
depend on two type parameters inheriting from \code{VariateForm}
and \code{ValueSupport}. The most generic form of such a construct
is represented by the \code{Sampleable} type, defining
something from which random samples can be drawn:
\begin{lstlisting}[language=Julia]
julia> abstract type Sampleable{F<:VariateForm,S<:ValueSupport}
end
\end{lstlisting}
A \code{Distribution} is a sub-type of \code{Sampleable},
carrying the same type parameter capturing the sample space:
\begin{lstlisting}[language=Julia]
julia> abstract type Distribution{F<:VariateForm,S<:ValueSupport} <: Sampleable{F,S}
end
\end{lstlisting}
A \code{Distribution} is more specific than a \code{Sampleable},
it describes the probability law mapping elements of a $\sigma$-algebra
(subsets of the sample space) to corresponding probabilities
of occurrence and is associated with corresponding probability
distribution functions (CDF, PDF).
As such, it extends the required interface as detailed in Section \ref{sec:distribution}.
In \pkg{Distributions.jl}, distribution families are represented
as types and particular distributions as instances of these types.
One advantage of this structure is the ease of defining a new
distribution family by creating a sub-type of \code{Distribution}
respecting the interface. The behavior of distributions
can also be extended by defining new functions over
all sub-types of \code{Distribution} and using the interface.


\section{Sampling interface}\label{sec:sampling}

Some programs require the generation of random values in a certain fashion
without requiring the analytical closed-form probability distribution.
The \code{Sampleable} type and interface serve these use cases.

A random quantity drawn from a sample space with given probability
distribution requires a way to sample values.
Such construct from which values can be sampled is
programmatically defined as an abstract type \code{Sampleable}
parameterized by \code{F} and \code{S}.
The first type parameter \code{F} classifies sampling objects and distribution
by their dimension, univariate distributions associated with scalar random variables,
multivariate distributions associated with vector random variables and matrix-variate
distributions associated with random matrices.
The second type parameter \code{S} specifies the support, discrete or continuous.
New value support and variate form types can also be defined,
subtyped from \code{ValueSupport} or \code{VariateForm}.

Furthermore, probability distributions are mathematical
functions to consider as immutable. A \code{Sampleable} on the other
hand can be a mutable object as shown below.

\begin{example}
Consider the following implementation of
a discrete N-state Markov chain as a \code{Sampleable}.
\begin{lstlisting}[language=Julia]
julia> using Distributions
julia> mutable struct MarkovChain <: Sampleable{Univariate,Discrete}
    s::Int
    m::Matrix{Float64}
end
\end{lstlisting}
A type implementation can only be a subtype of an abstract type
as \code{Sampleable}. With the structure defined, we can implement the required
method for a \code{Sampleable}, namely \code{rand} which is defined in \code{Base} Julia.
\begin{lstlisting}[language=Julia]
julia> import Base: rand
julia> import Random
julia> function rand(rng, mc::MarkovChain)
    r = rand(rng)
    v = cumsum(mc.m[mc.s,:])
    idx = findfirst(x -> x >= r, v)
    mc.s = idx
    return idx
end
julia> rand(mc::MarkovChain) = rand(Random.GLOBAL_RNG, mc)
\end{lstlisting}
\code{rng} is a random number generator (RNG) object, passing it as an
argument makes the \code{rand} implementation predictable.
If reproducibility is not important to the use case, \code{rand}
can be called as \code{rand(mc)} as implemented in the second method,
using the global random number generator \code{Random.GLOBAL_RNG}.
\end{example}

Note that the Markov chain implementation could have been defined in an
immutable way. This implementations however highlights one possible
definition of a \code{Sampleable} different from a probability
distribution.
This flexibility of implementation comes with a homogeneous interface,
any item from which random samples can be generated is called in the
same fashion. From the user perspective, where a sampler is defined
has no impact on its use thanks to \proglang{Julia} multiple dispatch
mechanism: the correct method (function implementation) is called
depending on the input type.

The \code{rand} function is the only required element for defining
the sampling with a particular process. Different methods are defined
for this function, specifying the pseudo-random number generator (PRNG)
to be used. The default RNG uses the  Mersenne-Twister algorithm
\citep{matsumoto1998mersenne}. The sampling state is kept in a global
variable used by default when no RNG is provided. New random number
generators can be defined by users, sub-typed from the \code{Random.AbstractRNG}
type.

\section{Distribution interface and types}\label{sec:distribution}

The core of the package are probability distributions, defined
as an abstract type as presented in \ref{sec:types}.
Any distribution must implement the \code{rand} method, which means random
values following the distribution can be generated. The two other essential
methods to implement are \code{pdf} giving the probability density function
at one point for continuous distributions or the probability mass function
at one point for discrete distributions and \code{cdf} evaluating the
Cumulated Density Function at one point. The \code{quantile} method from
the standard library \pkg{Statistics} module can be implemented for a
\code{Distribution} type, with the form
\code{quantile(d::Distribution,p::Number)} and returning the
value corresponding to the corresponding cumulative probability \code{p}.
Given that the method \code{rand()} without any argument follows
a uniform
pseudo-random number in the interval $\left[0,1\right]$, a default
fall-back method for random number generation can be defined
by inverse transform sampling for a univariate distribution as:
\begin{lstlisting}[language=Julia]
julia> rand(d::UnivariateDistribution) = quantile(d, rand())
\end{lstlisting}
The equivalent \proglang{R} functions for the normal distribution
can be matched to the \pkg{Distributions.jl} way of expressing them
as follows:
\begin{lstlisting}[language=Julia]
julia> using Distributions
julia> rnorm(n, mu, sig) = rand(Normal(mu, sig), n)
julia> dnorm(x, mu, sig) = pdf(Normal(mu, sig), x)
julia> pnorm(x, mu, sig) = cdf(Normal(mu, sig), x)
julia> qnorm(p, mu, sig) = quantile(Normal(mu, sig), p)
\end{lstlisting}
The advantage of using multiple dispatch is that supporting a
new distribution only requires the package API to grow
by one element which is the new distribution type,
instead of four new functions.
Most common probability distributions are defined by a mathematical
form and a set of parameters. All distributions must implement
the \code{params} method from the \pkg{StatsBase.jl}, allowing the
following example to always work for any given distribution type
\code{Dist} and \code{d} an instance of the distribution:
\begin{lstlisting}[language=Julia]
julia> p = params(d)
julia> new_dist = Dist(p...)
\end{lstlisting}
Depending on the distribution, sampling a batch of data can be done
in a more efficient manner than multiple individual samples.
Specialized samplers can be provided for distributions, letting user sample
batches of values.
Sampling from the Kolmogorov distribution for instance is done by pulling atomic
samples using an alternating series method \citep[IV.5]{devroye86}, while the Poisson distribution
lets users use either a counting sampler or Ahrens-Dieter sampler
\citep{ahrens1982computer}.

We will not expand explanations on most trivial functions of
the distribution interface, such as \code{minimum}, \code{maximum}
which can be defined in terms of \code{support}.
Other values are optional to define for distributions, such as
\code{mean}, \code{median}, \code{variance}. Not defining these
methods as mandatory allows for instance for the Cauchy-Lorentz
distribution to be defined.

\section{Distribution fitting and estimation}\label{sec:fitting}

Given a collection of samples and a distribution dependent on
a vector of parameters, the distribution fitting task consists
in finding an estimation of the distribution parameters $\hat{\theta}$.

\subsection{Maximum likelihood estimation}

Maximum Likelihood is a common technique for estimating the
parameters $\theta$ of a distribution given observations
\citep{wilks1938, aldrich1997ra}, with numerous applications in statistics
but also in signal processing \citep{mle_separation_sources}.
The \pkg{Distributions.jl} interface is defined as one
\code{fit_mle} function with two methods:
\begin{lstlisting}[language=Julia]
Distributions.fit_mle(D, x)
Distributions.fit_mle(D, x, w)
\end{lstlisting}
With \code{D} a distribution type to fit, \code{x} being
either a vector of observations if \code{D} is univariate,
a matrix of individuals/attributes if \code{D} is multivariate,
and \code{w} weights for the individual data points.
The function call returns an instance of \code{D} with the parameter
adjusted to the observations from \code{x}.

\begin{example}\label{exm:mle}
Fitting a normal distribution by maximum likelihood.
\begin{lstlisting}[language=Julia]
julia> using Distributions
julia> import Random
julia> Random.seed!(33)
julia> xs = rand(Normal(50.0, 10.0), 1000)
julia> fit_mle(Normal, xs)
Distributions.Normal{Float64}(µ=49.781, σ=10.002)
\end{lstlisting}
\end{example}

Additional partial information can be provided with
distribution-specific keywords.\linebreak
\code{fit_mle(Normal, x)} accepts for instance the keyword arguments
\code{mu} and \code{sigma} to specify either a fixed mean or standard
deviation.

The maximum likelihood estimation (MLE) function is implemented for
specific distribution types, corresponding to the fact that
there is no default and efficient MLE algorithm for a generic distribution.
The implementations present in \pkg{Distributions.jl} only require the
computation of sufficient statistics.
For a more advanced use cases, section \ref{sec:wineproduct} presents the
maximization of the likelihood for a custom distribution over a real
dataset with a Cartesian product of univariate distributions.
The behavior can be extended to a new distribution type \code{D} by
implementing \code{fit_mle} for it:
\begin{lstlisting}[language=Julia]
julia> using Distributions
julia> struct D <: ContinuousUnivariateDistribution end
julia> function Distributions.fit_mle(::Type{D}, xs::AbstractVector)
    # implementation
end
\end{lstlisting}

\subsection{Non-parametric estimation}\label{subsec:nonparam}

Some distribution estimation tasks have to be carried out without
the knowledge of the distribution form or family for various reasons.
Non-parametric estimation generally comprises the estimation of the
Cumulative Density Function and of the Probability Density Function.
\pkg{StatsBase.jl} provides the \code{ecdf} function, a higher
order function returning the empirical CDF at any new point.
Since the empirical CDF is built as a weighted sum of
Heaviside functions, it is discontinuous.

\pkg{StatsBase.jl} also provides a generic \code{fit} function,
used to build density histograms. \code{Histogram} is a type storing
the bins and heights, other functions from \pkg{Base} \proglang{Julia}
are defined on \code{Histogram} for convenience.
\code{fit} for histogram is defined with the following signature:
\begin{verbatim}
StatsBase.fit(Histogram, data[, weight][, edges]; closed=:right, nbins)
\end{verbatim}

\section{Modeling mixtures of distributions}\label{sec:mixtures}

Mixture models are used for modeling populations which consists of multiple sub-populations; mixture models are often applied to clustering or noise separation tasks.

A mixture distribution consists of several \emph{component distributions}, each of which has a relative \emph{component weight} or \emph{component prior probability}. For a mixture of $n$ components, then the densities can be written as a weighted sum:
\begin{equation*}
f_{\text{mix}}(x;\pi,\theta) = \sum_{i=1}^n \pi_i \, f(x,\theta_i)
\end{equation*}
where the parameters are the component weights $\pi = (\pi_1, \ldots, \pi_n)$,
taking values on the unit simplex, and each $\theta_i$ parameterizes the $i$th component distribution.

Sampling from a mixture model consists of first selecting the component according to the relative weight, then sampling from the corresponding component distribution. Therefore a mixture model can also be interpreted as a hierarchical model.

Mixtures are defined as an abstract sub-type of \code{Distribution},
parameterized by the same type information on the variate form and
value support:
\begin{lstlisting}[language=Julia]
julia> abstract type AbstractMixtureModel{VF<:VariateForm,VS<:ValueSupport}
<: Distribution{VF, VS}
end
\end{lstlisting}
Any \code{AbstractMixtureModel} is therefore a \code{Distribution} and
therefore implements specialized the mandatory methods \code{insupport},
\code{mean}, \code{var}, etc. Mixture models also need to implement the
following behavior:
\begin{itemize}
    \item \code{ncomponents(d::AbstractMixtureModel)}: returns the number of components in the mixture
    \item \code{component(d::AbstractMixtureModel, k)}: returns the k-th component
    \item \code{probs(d)}: returns the vector of prior probabilities over components
\end{itemize}

A concrete generic implementation is then defined as:
\begin{lstlisting}[language=Julia]
julia> struct MixtureModel{VF<:VariateForm,VS<:ValueSupport,Component<:Distribution}
<: AbstractMixtureModel{VF,VS}
    components::Vector{Component}
    prior::Categorical
end
\end{lstlisting}
Once constructed, it can be manipulated as any distribution
with the different methods of the interface.

\begin{example}
Figure \ref{fig:unigmm} shows the plot resulting from the following
construction of a univariate Gaussian mixture model.
\begin{lstlisting}[language=Julia]
julia> using Distributions
julia> import Plots
julia> gmm = MixtureModel(
    Normal.([-1.0, 0.0, 3.0],
            [0.3, 0.5, 1.0]),
    [0.25, 0.25, 0.5],
)
julia> xs = -2.0:0.01:6.0
julia> Plots.plot(xs, pdf.(gmm, xs), legend=nothing)
julia> Plots.ylabel!("\$f_X(x)\$")
julia> Plots.xlabel!("\$x\$")
julia> Plots.title!("Gaussian mixture PDF")
\end{lstlisting}
The \code{MixtureModel} constructor is taking as argument
a vector of \code{Normal} distributions, defined here from a vector
of mean values and a vector of standard deviation values.
\begin{figure}[t!]
    \centering
    \includegraphics[width = 0.6\textwidth]{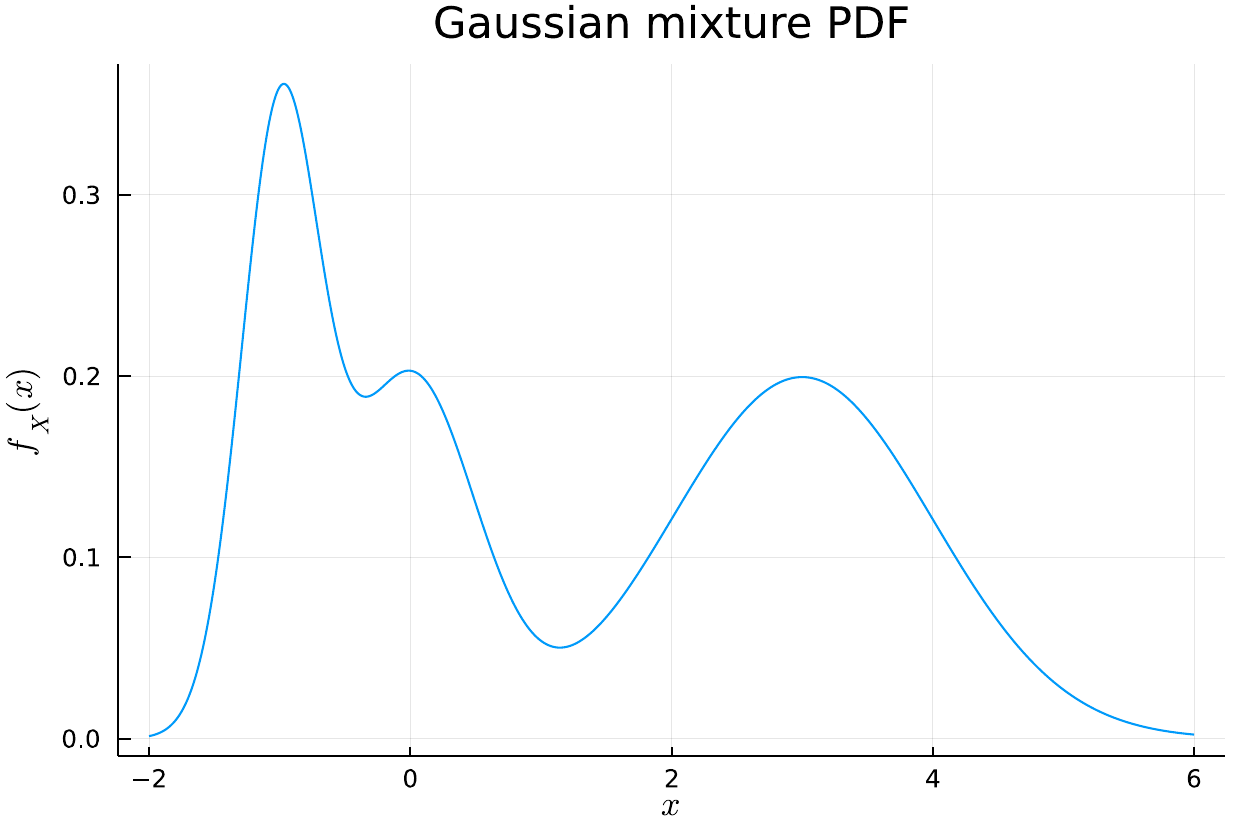}
    \caption{Example PDF of a univariate Gaussian mixture}
    \label{fig:unigmm}
\end{figure}
\end{example}
Mixture models remain an active area of research and of development within
\pkg{Distributions.jl}, estimation methods, and improved multivariate and
matrix-variate support are planned for future releases.
A more advanced example of estimation of a Gaussian
mixture by a generic EM algorithm is presented
in section \ref{sec:wineemalg}.

\section{Applications of Distributions.jl}\label{sec:applications}

\pkg{Distributions.jl} has become a foundation for statistical
programming in \proglang{Julia}, notably used for economic modeling
\citep{dsge, quantecon}, Markov chain Monte Carlo simulations \citep{mamba}
or randomized black-box optimization \cite{bboptim}
\footnote{Other packages depending on \pkg{Distributions.jl} can be found on the Julia General registry: \url{https://github.com/JuliaRegistries/General}}. We present below two applications of the package for non-parametric
continuous density estimation and probabilistic programming.

\subsection{Kernel density estimation}\label{sec:kde}

Probability density functions can be estimated in a non-parametric
fashion using kernel density estimation  \citep[KDE,][]{rosenblatt1956,parzen1962}.
This is supported through the \pkg{KernelDensity.jl}
package, defining the \code{kde} function to infer an estimate density from
data. Both univariate and bivariate density estimates are supported.
Most of the algorithms and parameter selection heuristics developed
in \pkg{KernelDensity.jl} are based on \cite{silverman2018density}.
\pkg{KernelDensity.jl} supports multiple distributions as base kernels, and
can be extended to other families. The default kernel used is Gaussian.

\begin{figure}[t!]
    \centering
    \includegraphics[width = 0.6\textwidth]{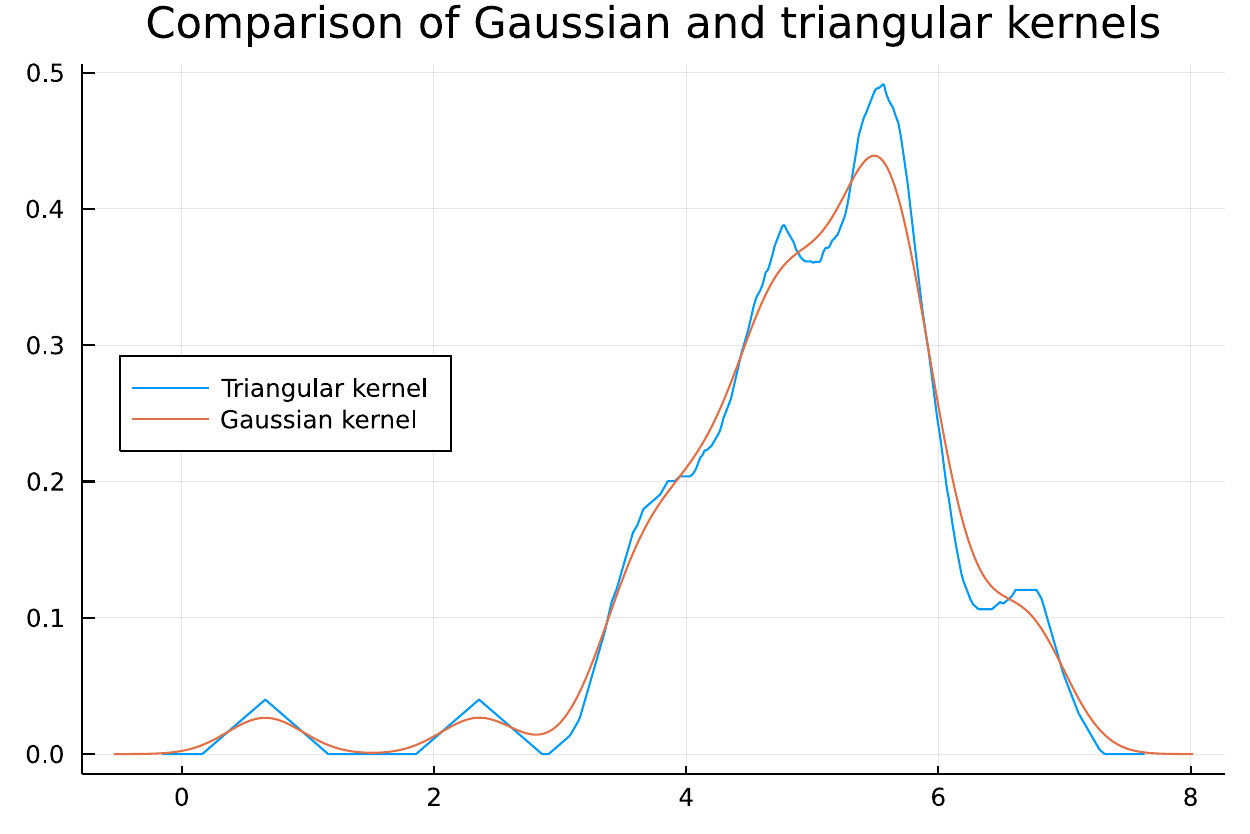}
    \caption{Comparison of different kernels defined as Distribution objects}
    \label{fig:kde_compare2}
\end{figure}

\begin{example}
We highlight the estimation of a kernel density on data generated from the
mixture of a log-normal and uniform distributions.
\begin{lstlisting}[language=Julia]
julia> import Random, Plots, KernelDensity
julia> using Distributions
julia> function generate_point(rng = Random.GLOBAL_RNG)
    thres = rand(rng)
    if thres >= 0.5
        rand(rng, LogNormal())
    else
        rand(rng, Uniform(2.0, 3.0))
    end
end
julia> mt = Random.MersenneTwister(42)
julia> xs = [generate_point(mt) for _ in 1:5000]
julia> bandwidths = [0.05, 0.1, 0.5]
julia> densities = [KernelDensity.kde(xs, bandwidth = bw) for bw in bandwidths]
julia> p = Plots.plot()
julia> for (b,d) in zip(bandwidths, densities)
    Plots.plot!(p, d.x, d.density, labels = "bw = $b")
end
julia> Plots.xlims!(p, 0.0, 8.0)
julia> Plots.title!("KDE with Gaussian Kernel")
\end{lstlisting}
\begin{figure}[t!]
    \centering
    \includegraphics[width = 0.6\textwidth]{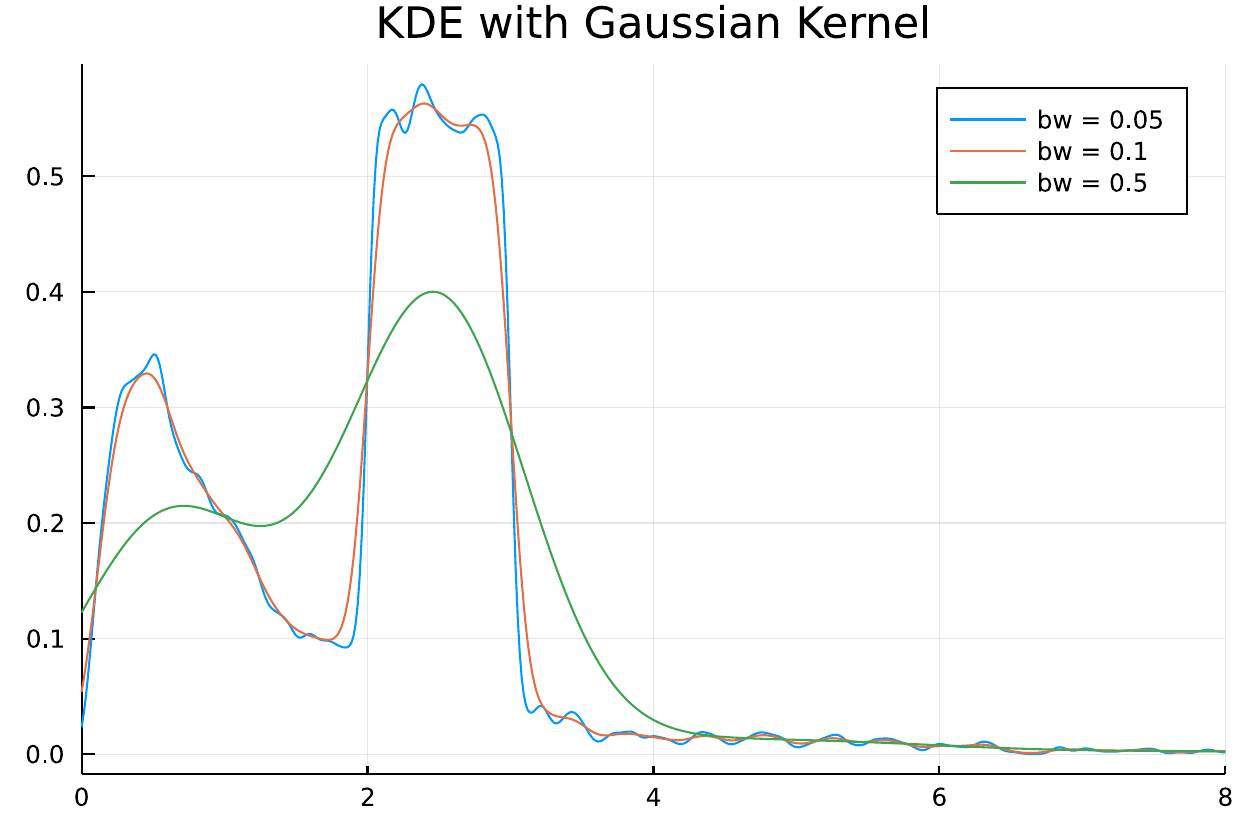}
    \caption{Adjusted KDE with various bandwidths}
    \label{fig:kde}
\end{figure}
The bandwidth $bw = 0.1$ seems not to overfit isolated data points
without smoothing out important components.
All examples provided use the \pkg{Plots.jl}
\footnote{http://docs.juliaplots.org} package to plot results,
which can be used with various plotting engines as back-end.
We can compare the kernel density estimate to the real PDF
as done in the following script and illustrated Figure
\ref{fig:kde_compare}.
\begin{lstlisting}[language=Julia]
julia> xvals = 0.01:0.01:8.0
julia> yvals = map(xvals) do x
    comp1 = pdf(LogNormal(), x)
    comp2 = pdf(Uniform(2.0, 3.0), x)
    0.5 * comp1 + 0.5 * comp2
end
julia> p = Plots.plot(xvals, yvals, labels = "Real distribution")
julia> kde = KernelDensity.kde(xs, bandwidth = 0.1)
julia> Plots.plot!(p, kde.x, kde.density, labels = "KDE")
julia> Plots.plot!(p, xvals, yvals, labels = "Real distribution")
julia> Plots.xlims!(p, 0.0, 8.0)
\end{lstlisting}
\begin{figure}[t!]
    \centering
    \includegraphics[width = 0.6\textwidth]{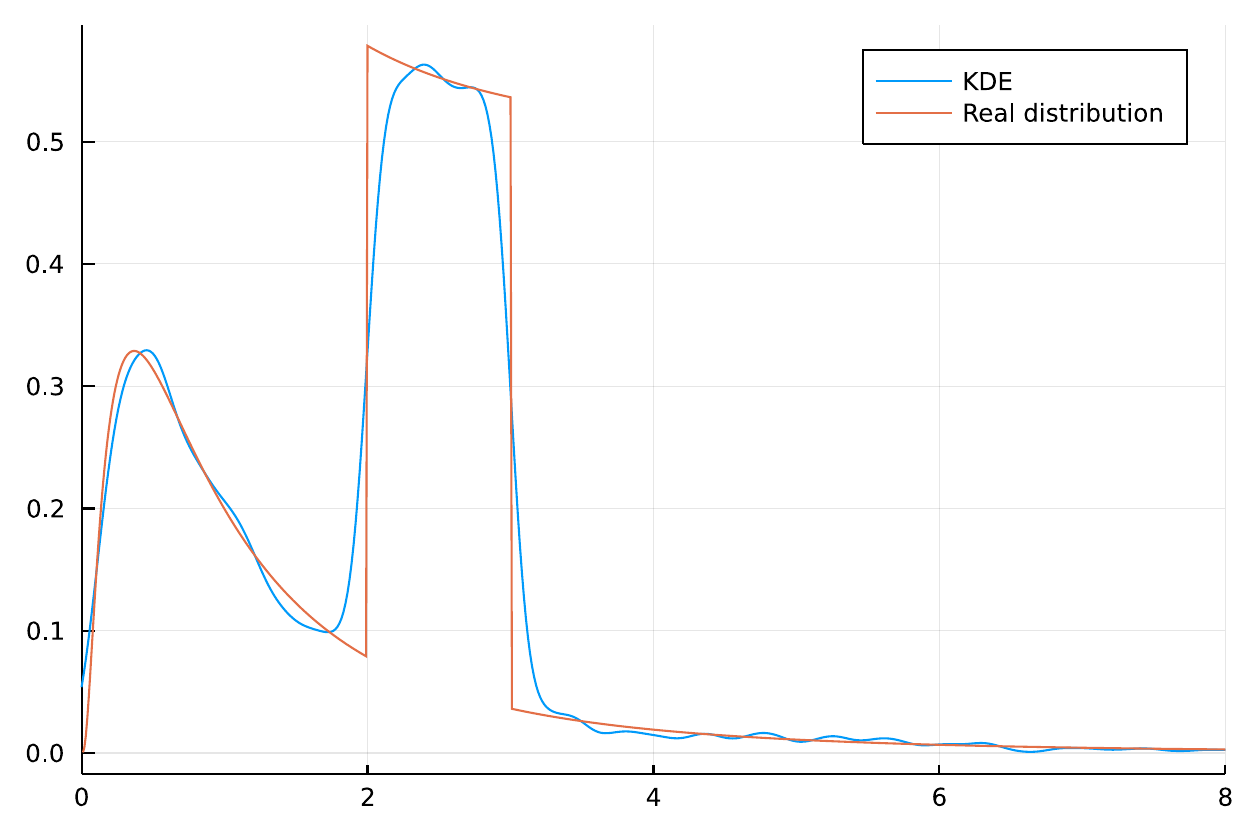}
    \caption{Comparison of the experimental and real PDF}
    \label{fig:kde_compare}
\end{figure}
\end{example}
The kernel density estimation technique relies on a base distribution
(the kernel) which is convoluted against the data.
The package uses directly the interface presented in section
\ref{sec:distribution} to accept a \code{Distribution} as second
parameter. The following script computes the kernel density estimates
with a Gaussian and triangular distributions. The result is illustrated
Figure \ref{fig:kde_compare2}.
\begin{lstlisting}[language=Julia]
julia> mt = Random.MersenneTwister(42)
julia> (µ, σ) = (5.0, 1.0)
julia> xs = [rand(mt, Normal(µ, σ)) for _ in 1:50]
julia> ndist = Normal(0.0, 0.3)
julia> gkernel = kde(xs, ndist)
julia> tdist = TriangularDist(-0.5, 0.5)
julia> tkernel = KernelDensity.kde(xs, tdist)
julia> p = Plots.plot(tkernel.x, tkernel.density, labels = "Triangular kernel")
julia> Plots.plot!(p, gkernel.x, gkernel.density,
            labels = "Gaussian kernel", legend = :left)
julia> Plots.title!(p, "Comparison of Gaussian and triangular kernels")
\end{lstlisting}
The density estimator is computed via the Fourier transform: any distribution with a
defined characteristic function (via the \code{cf} function from
\pkg{Distributions.jl}) can be used as a kernel.
The ability to manipulate distributions as types and objects
allows end-users and package developers to compose on top
of defined distributions and to define their own to use
for kernel density estimations without any additional runtime cost.
The kernel estimation of a bi-variate density is shown in \ref{sec:winekde}.

\subsection{Probabilistic programming languages}\label{sec:ppl}

Probabilistic programs are computer programs with the added capability of
drawing the value of a variable from a probability distribution and conditioning variable
values on observations \citep{gordon2014probabilistic}.
They allow for the specification of complex relations between random variables
using a set of constructs that go beyond typical graphical models to include flow
control (e.g., loops, conditions), recursion, user-defined types and other algorithmic building blocks,
differing from explicit graph construction by users, as done for instance in \cite{mamba}.
Probabilistic programming languages (PPL) are programming language or libraries
enabling developers to write such a probabilistic program. That is, they are a type of
domain-specific language \citep[DSL,][]{fowler2010domain}.
We refer the interested reader to \cite{van2018introduction} for an overview of
PPL design and use, to \cite{scibior2017denotational} for the analysis of
Bayesian inference algorithms as transformations of intermediate representations,
and to \cite{vakar2019domain} for representation of recursive types in PPLs.
They often fall in two categories.

In the first type, the PPL is its own independent language, including
syntax and parsing, though it may rely on another ``host'' language for its execution engine.
For example, \proglang{Stan} \citep{stan}, uses its own \code{.stan} model
specification format, though the parser and inference engine are written in \proglang{C++}.
This type of approach benefits from defining its own syntax, thus designing it to
look similar to the way equivalent statistical models are written. The model is
also verified for syntactic correctness at compile-time.\footnote{
An example model combining \pkg{Stan} with \proglang{R} can be found
at \cite{golf}.
}

In the second type of PPL, the language is embedded within and makes use of the
syntax of the host language, leveraging that language's native constructs.
For example, \pkg{PyMC} \citep{pymc, kochurov2019pymc4}, Pyro \citep{pyro}, and \pkg{Edward} \citep{edward} all
use \proglang{Python} as the host language, leveraging \pkg{Theano}, \pkg{Torch}, and \pkg{TensorFlow},
respectively, to perform many of the underlying inference computations. Here, the
key advantage is that these PPLs gain access to the ecosystem of the host language
(data structures, libraries, tooling). User documentation and development efforts
can also focus on the key aspects of a PPL instead of the full stack.\footnote{
The same golf putting data were used using \pkg{PyMC} as PPL in \cite{golfpymc}.
}

However, both approaches to PPLs suffer from drawbacks. In the first type of PPLs,
users must add new elements to their toolchains for the development and inference
of statistical models. That is, they are likely to use a general-purpose
programming language for other tasks, then switch to the PPL environment for the
sampling and inference task, and then read the result back into the general-purpose
programming language. One solution to this problem is to develop APIs in general
purpose languages, but full inter-operability between two environments is non-trivial.
Developers must then build a whole ecosystem around the PPL, including data
structures, input/output, and text editor support, which costs time that might
otherwise go toward improving the inference algorithms and optimization procedures.

While the second type of PPL may not require the development of a separate,
parallel set of tools, it often runs into ``impedance mismatches'' of its own.
In many cases, the host language has not been designed for a re-use of its constructs
for probabilistic programming, resulting in syntax that aligns poorly with users'
mathematical intuitions. Moreover, duplication may still occur at the level of
classes or libraries, as, for example, Pyro and Edward depend on Torch and TensorFlow,
which replicate much of the linear algebra functionality of NumPy for their own
array constructs.

By contrast, the design of \pkg{Distributions.jl} has enabled the development
of embedded probabilistic programming languages such as \pkg{Turing.jl}
\citep{DBLP:conf/aistats/GeXG18}, \pkg{SOSS.jl} \citep{soss}, \pkg{Omega.jl} \citep{omega}
or \pkg{Poirot.jl} \citep{poirot}
with comparatively less overhead or friction with the host language. These PPLs
are able to make use of three elements unique to the \proglang{Julia} ecosystem
to challenge the dichotomy between embedded and stand-alone PPLs.
First, they make use of \proglang{Julia}'s rich type system and multiple dispatch,
which more easily support modeling mathematical constructs as types.
Second, they all utilize \pkg{Distributions.jl}'s types and hierarchy for the
sampling of random values and representation of their distributions.
Finally, they use manipulations and transformations of the user \proglang{Julia}
code to create new syntax that matches domain-specific requirements.

These transformations are either performed through macros
(for \pkg{Soss.jl} and \pkg{Turing.jl}) or modifications of the compilation
phases through compiler overdubbing (\pkg{Omega.jl} and \pkg{Poirot.jl} use
\pkg{Cassette.jl} and \pkg{IRTools.jl} respectively to modify the user code or
its intermediate representations).
As in the \proglang{Lisp} tradition, \proglang{Julia}'s macros are written and
manipulable in \proglang{Julia} itself and rewrite code during the lowering phase \citep{Bezanson2018}.
Macros allow PPL designers to keep programs close to standard \proglang{Julia} while introducing package-specific
syntax that more closely mimics statistical conventions, all without compromising on performance.
For instance, a simple model definition from the \pkg{Turing.jl} documentation illustrates a model that
is both readable as \proglang{Julia} code while staying close to statistical conventions:
\begin{lstlisting}[language=Julia]
julia> @model coinflip(y) = begin
    p ~ Beta(1, 1)

    N = length(y)
    for n in 1:N
        y[n] ~ Bernoulli(p)
    end
end
\end{lstlisting}
With these syntax manipulation capabilities, along with compiled language performance,
the combination of \proglang{Julia} with \pkg{Distributions.jl} provides an
excellent foundation for further research and development of PPLs.
\footnote{
The golf putting model was ported to \pkg{Turing.jl} in \cite{golfturing}.
}

\section{Conclusion and future work}\label{sec:conclusion}

We presented some of the types, structures and tools for
modeling and computing on probability distributions in the
\pkg{JuliaStats} ecosystem.
The \pkg{JuliaStats} packages leverage some key constructs of
\proglang{Julia} such as definition of new composite types,
abstract parametric types and sub-typing, along with multiple dispatch.
This allows users to express computations involving random
number generation and the manipulation of probability distributions.
A clear interface allows package developers to build new distributions
from their mathematical definition. New algorithms can be developed,
using the Distribution interface and as such extending
the new features to all sub-types.
These two features of \proglang{Julia}, namely extension of behavior to new types
and definition of behavior over existing type hierarchy,
allow different features built around the Distribution type
to inter-operate seamlessly without hard coupling nor additional
work from the package developers or end-users.
Future work on the package will include the implementation of
maximum likelihood estimation for other distributions,
including mixtures and matrix-variate distributions.

\section*{Acknowledgments}

The authors thank and acknowledge the work of all contributors
and maintainers of packages of the \proglang{JuliaStats} ecosystem
and especially Dave Kleinschmidt for his valuable feedback
on early versions of this article, Andreas Noack for the overall
supervision of the package, including the integration of numerous
pull requests from external contributors, Moritz Schauer
for discussions on the representability of mathematical objects
in the \proglang{Julia} type system and Zenna Tavares for the critical
review of early drafts.

Research sponsored by the Laboratory Directed Research and Development Program
of Oak Ridge National Laboratory, managed by UT-Battelle, LLC, for the US
Department of Energy under contract DE-AC05-00OR22725. This research used
resources of the Compute and Data Environment for Science (CADES) at the Oak
Ridge National Laboratory, which is supported by the Office of Science of the
U.S. Department of Energy under Contract No.~DE-AC05-00OR22725. This applies
specifically to funding provided by the Oak Ridge National Laboratory to
Theodore Papamarkou.

\bibliography{v98i16}

\begin{thebibliography}{47}
\newcommand{\enquote}[1]{``#1''}
\providecommand{\natexlab}[1]{#1}
\providecommand{\url}[1]{\texttt{#1}}
\providecommand{\urlprefix}{URL }
\expandafter\ifx\csname urlstyle\endcsname\relax
  \providecommand{\doi}[1]{doi:\discretionary{}{}{}#1}\else
  \providecommand{\doi}{doi:\discretionary{}{}{}\begingroup
  \urlstyle{rm}\Url}\fi
\providecommand{\eprint}[2][]{\url{#2}}

\bibitem[{Aeberhard \emph{et~al.}(1994)Aeberhard, Coomans, and {De
  Vel}}]{aeberhard1994comparative}
Aeberhard S, Coomans D, {De Vel} O (1994).
\newblock \enquote{Comparative Analysis of Statistical Pattern Recognition
  Methods in High Dimensional Settings.}
\newblock \emph{Pattern Recognition}, \textbf{27}(8), 1065--1077.
\newblock \doi{10.1016/0031-3203(94)90145-7}.

\bibitem[{Ahrens and Dieter(1982)}]{ahrens1982computer}
Ahrens JH, Dieter U (1982).
\newblock \enquote{Computer Generation of Poisson Deviates from Modified Normal
  Distributions.}
\newblock \emph{ACM Transactions on Mathematical Software}, \textbf{8}(2),
  163--179.
\newblock \doi{10.1145/355993.355997}.

\bibitem[{Aldrich(1997)}]{aldrich1997ra}
Aldrich J (1997).
\newblock \enquote{R.A.~Fisher and the Making of Maximum Likelihood
  1912--1922.}
\newblock \emph{Statistical Science}, \textbf{12}(3), 162--176.
\newblock \doi{10.1214/ss/1030037906}.

\bibitem[{Bezanson \emph{et~al.}(2018)Bezanson, Chen, Chung, Karpinski, Shah,
  Vitek, and Zoubritzky}]{Bezanson2018}
Bezanson J, Chen J, Chung B, Karpinski S, Shah VB, Vitek J, Zoubritzky L
  (2018).
\newblock \enquote{\proglang{Julia}: Dynamism and Performance Reconciled by
  Design.}
\newblock \emph{Proceedings of the ACM on Programming Languages},
  \textbf{2}(OOPSLA), 120.
\newblock \doi{10.1145/3276490}.

\bibitem[{Bezanson \emph{et~al.}(2017)Bezanson, Edelman, Karpinski, and
  Shah}]{julia101}
Bezanson J, Edelman A, Karpinski S, Shah V (2017).
\newblock \enquote{\proglang{Julia}: A Fresh Approach to Numerical Computing.}
\newblock \emph{SIAM Review}, \textbf{59}(1), 65--98.
\newblock \doi{10.1137/141000671}.

\bibitem[{Bingham \emph{et~al.}(2019)Bingham, Chen, Jankowiak, Obermeyer,
  Pradhan, Karaletsos, Singh, Szerlip, Horsfall, and Goodman}]{pyro}
Bingham E, Chen JP, Jankowiak M, Obermeyer F, Pradhan N, Karaletsos T, Singh R,
  Szerlip P, Horsfall P, Goodman ND (2019).
\newblock \enquote{\pkg{Pyro}: Deep Universal Probabilistic Programming.}
\newblock \emph{The Journal of Machine Learning Research}, \textbf{20}(1),
  973--978.

\bibitem[{{\pkg{Boost} Developers}(2018)}]{boost}
{\pkg{Boost} Developers} (2018).
\newblock \enquote{{\pkg{Boost} \proglang{C++} Libraries}.}
\newblock \urlprefix\url{https://www.boost.org/}.

\bibitem[{Carpenter \emph{et~al.}(2017)Carpenter, Gelman, Hoffman, Lee,
  Goodrich, Betancourt, Brubaker, Guo, Li, and Riddell}]{stan}
Carpenter B, Gelman A, Hoffman MD, Lee D, Goodrich B, Betancourt M, Brubaker M,
  Guo J, Li P, Riddell A (2017).
\newblock \enquote{\proglang{Stan}: A Probabilistic Programming Language.}
\newblock \emph{Journal of Statistical Software}, \textbf{76}(1), 1--32.
\newblock \doi{10.18637/jss.v076.i01}.

\bibitem[{Devroye(1986)}]{devroye86}
Devroye L (1986).
\newblock \emph{Non-Uniform Random Variate Generation}.
\newblock Springer-Verlag, New York.

\bibitem[{Duncan(2020)}]{golfturing}
Duncan J (2020).
\newblock \enquote{Model building of golf putting with \proglang{Turing.jl}.}
\newblock
  \urlprefix\url{https://web.archive.org/web/20210611145623/https://jduncstats.com/posts/2019-11-02-golf-turing/}.

\bibitem[{Erwig and Kollmansberger(2006)}]{haskell}
Erwig M, Kollmansberger S (2006).
\newblock \enquote{Functional Pearls: Probabilistic Functional Programming in
  \proglang{Haskell}.}
\newblock \emph{Journal of Functional Programming}, \textbf{16}(1), 21--34.
\newblock \doi{10.1017/s0956796805005721}.

\bibitem[{{Federal Reserve Bank Of New York}(2019)}]{dsge}
{Federal Reserve Bank Of New York} (2019).
\newblock \enquote{\pkg{DSGE.jl}: Solve and Estimate Dynamic Stochastic General
  Equilibrium Models (Including the New York Fed DSGE).}
\newblock \urlprefix\url{https://github.com/FRBNY-DSGE/DSGE.jl}.

\bibitem[{Feldt(2017)}]{bboptim}
Feldt R (2017).
\newblock \enquote{\pkg{BlackBoxOptim.jl}.}
\newblock \urlprefix\url{https://github.com/robertfeldt/BlackBoxOptim.jl}.

\bibitem[{Fowler(2010)}]{fowler2010domain}
Fowler M (2010).
\newblock \emph{Domain-Specific Languages}.
\newblock Pearson Education.

\bibitem[{Ge \emph{et~al.}(2018)Ge, Xu, and
  Ghahramani}]{DBLP:conf/aistats/GeXG18}
Ge H, Xu K, Ghahramani Z (2018).
\newblock \enquote{\proglang{Turing}: Composable Inference for Probabilistic
  Programming.}
\newblock In \emph{Proceedings of the Twenty-First International Conference on
  Artificial Intelligence and Statistics}, pp. 1682--1690.
\newblock \urlprefix\url{http://proceedings.mlr.press/v84/ge18b.html}.

\bibitem[{Gelman(2019)}]{golf}
Gelman A (2019).
\newblock \enquote{Model Building and Expansion for Golf Putting (accessed June
  2021).}
\newblock
  \urlprefix\url{https://mc-stan.org/users/documentation/case-studies/golf.html}.

\bibitem[{Gordon \emph{et~al.}(2014)Gordon, Henzinger, Nori, and
  Rajamani}]{gordon2014probabilistic}
Gordon AD, Henzinger TA, Nori AV, Rajamani SK (2014).
\newblock \enquote{Probabilistic Programming.}
\newblock In \emph{Proceedings of the on Future of Software Engineering}, pp.
  167--181. ACM.

\bibitem[{Innes(2019)}]{poirot}
Innes M (2019).
\newblock \enquote{\pkg{Poirot.jl}.}
\newblock Version~0.1, \urlprefix\url{https://github.com/MikeInnes/Poirot.jl}.

\bibitem[{{JuliaStats}(2019)}]{distributions}
{JuliaStats} (2019).
\newblock \enquote{\pkg{Distributions.jl}.}
\newblock \doi{10.5281/zenodo.2647458}.

\bibitem[{Kiselyov and Shan(2009)}]{funcprobprog}
Kiselyov O, Shan CC (2009).
\newblock \enquote{Embedded Probabilistic Programming.}
\newblock In WM~Taha (ed.), \emph{Domain-Specific Languages}, pp. 360--384.
  Springer-Verlag, Berlin.

\bibitem[{Kochurov \emph{et~al.}(2019)Kochurov, Carroll, Wiecki, and
  Lao}]{kochurov2019pymc4}
Kochurov M, Carroll C, Wiecki T, Lao J (2019).
\newblock \enquote{\pkg{PyMC4}: Exploiting Coroutines for Implementing a
  Probabilistic Programming Framework.}
\newblock \urlprefix\url{https://openreview.net/forum?id=rkgzj5Za8H}.

\bibitem[{{Le Goc} and Donz{\'e}(2015)}]{LEGOC2015531}
{Le Goc} Y, Donz{\'e} A (2015).
\newblock \enquote{\pkg{EVL}: A Framework for Multi-Methods in \proglang{C++}.}
\newblock \emph{Science of Computer Programming}, \textbf{98}, 531--550.
\newblock ISSN 0167-6423.
\newblock \doi{10.1016/j.scico.2014.08.003}.

\bibitem[{Lyon \emph{et~al.}(2017)Lyon, Sargent, and Stachurski}]{quantecon}
Lyon S, Sargent TJ, Stachurski J (2017).
\newblock \enquote{\pkg{QuantEcon.jl}.}
\newblock \urlprefix\url{https://github.com/QuantEcon/QuantEcon.jl}.

\bibitem[{Matsumoto and Nishimura(1998)}]{matsumoto1998mersenne}
Matsumoto M, Nishimura T (1998).
\newblock \enquote{Mersenne Twister: A 623-Dimensionally Equidistributed
  Uniform Pseudo-Random Number Generator.}
\newblock \emph{ACM Transactions on Modeling and Computer Simulation},
  \textbf{8}(1), 3--30.
\newblock \doi{10.1145/272991.272995}.

\bibitem[{Mogensen and Riseth(2018)}]{mogensen2018optim}
Mogensen PK, Riseth AN (2018).
\newblock \enquote{\pkg{Optim}: A Mathematical Optimization Package for
  \proglang{Julia}.}
\newblock \emph{Journal of Open Source Software}, \textbf{3}(24), 615.
\newblock \doi{10.21105/joss.00615}.

\bibitem[{Orban and Siqueira(2019)}]{jso-2019}
Orban D, Siqueira AS (2019).
\newblock \enquote{\pkg{JuliaSmoothOptimizers}: Infrastructure and Solvers for
  Continuous Optimization in \proglang{Julia}.}
\newblock \doi{10.5281/zenodo.2655082}.

\bibitem[{Parzen(1962)}]{parzen1962}
Parzen E (1962).
\newblock \enquote{On Estimation of a Probability Density Function and Mode.}
\newblock \emph{The Annals of Mathematical Statistics}, \textbf{33}(3),
  1065--1076.
\newblock \doi{10.1214/aoms/1177704472}.

\bibitem[{Patil \emph{et~al.}(2010)Patil, Huard, and Fonnesbeck}]{pymc}
Patil A, Huard D, Fonnesbeck C (2010).
\newblock \enquote{\pkg{PyMC}: Bayesian Stochastic Modelling in
  \proglang{Python}.}
\newblock \emph{Journal of Statistical Software}, \textbf{35}(4), 1--81.
\newblock \doi{10.18637/jss.v035.i04}.

\bibitem[{Pham and Garat(1997)}]{mle_separation_sources}
Pham DT, Garat P (1997).
\newblock \enquote{Blind Separation of Mixture of Independent Sources through a
  Quasi-Maximum Likelihood Approach.}
\newblock \emph{IEEE Transactions on Signal Processing}, \textbf{45}(7),
  1712--1725.
\newblock \doi{10.1109/78.599941}.

\bibitem[{Pirkelbauer \emph{et~al.}(2010)Pirkelbauer, Solodkyy, and
  Stroustrup}]{cppdispatch}
Pirkelbauer P, Solodkyy Y, Stroustrup B (2010).
\newblock \enquote{Design and Evaluation of \proglang{C++} Open Multi-Methods.}
\newblock \emph{Science of Computer Programming}, \textbf{75}(7), 638--667.
\newblock \doi{10.1016/j.scico.2009.06.002}.
\newblock Generative Programming and Component Engineering (GPCE 2007).

\bibitem[{\pkg{PyMC}(2019)}]{golfpymc}
\pkg{PyMC} (2019).
\newblock \enquote{Model building and expansion for golf putting.}
\newblock
  \urlprefix\url{http://web.archive.org/web/20191210110708/https://nbviewer.jupyter.org/github/pymc-devs/pymc3/blob/master/docs/source/notebooks/putting_workflow.ipynb}.

\bibitem[{Revels \emph{et~al.}(2016)Revels, Lubin, and
  Papamarkou}]{revels2016forward}
Revels J, Lubin M, Papamarkou T (2016).
\newblock \enquote{Forward-Mode Automatic Differentiation in \proglang{Julia}.}
\newblock \emph{arXiv 1607.07892}, arXiv.org E-Print Archive.
\newblock \urlprefix\url{https://arxiv.org/abs/1607.07892}.

\bibitem[{Rosenblatt(1956)}]{rosenblatt1956}
Rosenblatt M (1956).
\newblock \enquote{Remarks on Some Nonparametric Estimates of a Density
  Function.}
\newblock \emph{The Annals of Mathematical Statistics}, \textbf{27}(3),
  832--837.
\newblock \doi{10.1214/aoms/1177728190}.

\bibitem[{Ruckdeschel \emph{et~al.}(2006)Ruckdeschel, Kohl, Stabla, and
  Camphausen}]{distr}
Ruckdeschel P, Kohl M, Stabla T, Camphausen F (2006).
\newblock \enquote{\proglang{S}4 Classes for Distributions.}
\newblock \emph{\proglang{R} News}, \textbf{6}(2), 2--6.

\bibitem[{Schauer(2018)}]{bridge}
Schauer M (2018).
\newblock \enquote{\pkg{Bridge.jl}: a Statistical Toolbox for Diffusion
  Processes and Stochastic Differential Equations.}
\newblock \doi{10.5281/zenodo.891230}.
\newblock Version~0.9.

\bibitem[{Scherrer(2019)}]{soss}
Scherrer C (2019).
\newblock \enquote{\pkg{Soss.jl}: Probabilistic Programming via Source
  Rewriting.}
\newblock Version~0.1, \urlprefix\url{https://github.com/cscherrer/Soss.jl}.

\bibitem[{{\'S}cibior \emph{et~al.}(2015){\'S}cibior, Ghahramani, and
  Gordon}]{scibior2015practical}
{\'S}cibior A, Ghahramani Z, Gordon AD (2015).
\newblock \enquote{Practical Probabilistic Programming with Monads.}
\newblock \emph{ACM SIGPLAN Notices}, \textbf{50}(12), 165--176.
\newblock \doi{10.1145/2887747.2804317}.

\bibitem[{{\'S}cibior \emph{et~al.}(2017){\'S}cibior, Kammar, V{\'a}k{\'a}r,
  Staton, Yang, Cai, Ostermann, Moss, Heunen, and
  Ghahramani}]{scibior2017denotational}
{\'S}cibior A, Kammar O, V{\'a}k{\'a}r M, Staton S, Yang H, Cai Y, Ostermann K,
  Moss SK, Heunen C, Ghahramani Z (2017).
\newblock \enquote{Denotational Validation of Higher-Order Bayesian Inference.}
\newblock \emph{Proceedings of the ACM on Programming Languages}, \textbf{2},
  60.
\newblock \doi{10.1145/3158148}.

\bibitem[{Silverman(2018)}]{silverman2018density}
Silverman BW (2018).
\newblock \emph{Density Estimation for Statistics and Data Analysis}.
\newblock Routledge.

\bibitem[{Smith(2018)}]{mamba}
Smith BJ (2018).
\newblock \enquote{\pkg{Mamba.jl}: Markov Chain Monte Carlo (MCMC) for Bayesian
  Analysis in \proglang{Julia}.}
\newblock Version~0.12,
  \urlprefix\url{https://github.com/brian-j-smith/Mamba.jl}.

\bibitem[{Tavares(2018)}]{omega}
Tavares Z (2018).
\newblock \enquote{\pkg{Omega.jl}: Causal, Higher-Order, Probabilistic
  Programming.}
\newblock Version~0.1, \urlprefix\url{https://github.com/zenna/Omega.jl}.

\bibitem[{Tran \emph{et~al.}(2016)Tran, Kucukelbir, Dieng, Rudolph, Liang, and
  Blei}]{edward}
Tran D, Kucukelbir A, Dieng AB, Rudolph M, Liang D, Blei DM (2016).
\newblock \enquote{\pkg{Edward}: A Library for Probabilistic Modeling,
  Inference, and Criticism.}
\newblock \emph{arXiv 1610.09787}, arXiv.org E-Print Archive.
\newblock \urlprefix\url{https://arxiv.org/abs/1610.09787}.

\bibitem[{V{\'a}k{\'a}r \emph{et~al.}(2019)V{\'a}k{\'a}r, Kammar, and
  Staton}]{vakar2019domain}
V{\'a}k{\'a}r M, Kammar O, Staton S (2019).
\newblock \enquote{A Domain Theory for Statistical Probabilistic Programming.}
\newblock \emph{Proceedings of the ACM on Programming Languages},
  \textbf{3}(POPL), 36.
\newblock \doi{10.1145/3290349}.

\bibitem[{{Van de Meent} \emph{et~al.}(2018){Van de Meent}, Paige, Yang, and
  Wood}]{van2018introduction}
{Van de Meent} JW, Paige B, Yang H, Wood F (2018).
\newblock \enquote{An Introduction to Probabilistic Programming.}
\newblock \emph{arXiv 1809.10756}, arXiv.org E-Print Archive.
\newblock \urlprefix\url{https://arxiv.org/abs/1809.10756}.

\bibitem[{Virtanen \emph{et~al.}(2020)Virtanen, Gommers, Oliphant, Haberland,
  Reddy, Cournapeau, Burovski, Peterson, Weckesser, Bright, {van der Walt},
  Brett, Wilson, Millman, Mayorov, Nelson, Jones, Kern, Larson, Carey, Polat,
  Feng, Moore, {VanderPlas}, Laxalde, Perktold, Cimrman, Henriksen, Quintero,
  Harris, Archibald, Ribeiro, Pedregosa, {van Mulbregt}, and {SciPy 1.0
  Contributors}}]{scipy}
Virtanen P, Gommers R, Oliphant TE, Haberland M, Reddy T, Cournapeau D,
  Burovski E, Peterson P, Weckesser W, Bright J, {van der Walt} SJ, Brett M,
  Wilson J, Millman KJ, Mayorov N, Nelson ARJ, Jones E, Kern R, Larson E, Carey
  CJ, Polat {\.I}, Feng Y, Moore EW, {VanderPlas} J, Laxalde D, Perktold J,
  Cimrman R, Henriksen I, Quintero EA, Harris CR, Archibald AM, Ribeiro AH,
  Pedregosa F, {van Mulbregt} P, {SciPy 10 Contributors} (2020).
\newblock \enquote{\pkg{SciPy} 1.0: Fundamental Algorithms for Scientific
  Computing in \proglang{Python}.}
\newblock \emph{Nature Methods}, \textbf{17}, 261--272.
\newblock \doi{10.1038/s41592-019-0686-2}.

\bibitem[{Wilks(1938)}]{wilks1938}
Wilks SS (1938).
\newblock \enquote{The Large-Sample Distribution of the Likelihood Ratio for
  Testing Composite Hypotheses.}
\newblock \emph{The Annals of Mathematical Statistics}, \textbf{9}(1), 60--62.
\newblock \doi{10.1214/aoms/1177732360}.

\bibitem[{{Zappa Nardelli} \emph{et~al.}(2018){Zappa Nardelli}, Belyakova,
  Pelenitsyn, Chung, Bezanson, and Vitek}]{ZappaNardelli18}
{Zappa Nardelli} F, Belyakova J, Pelenitsyn A, Chung B, Bezanson J, Vitek J
  (2018).
\newblock \enquote{\proglang{Julia} Subtyping: A Rational Reconstruction.}
\newblock \emph{Proceedings of the ACM on Programming Languages},
  \textbf{2}(OOPSLA), 113.
\newblock \doi{10.1145/3276483}.

\end{thebibliography}

\newpage

\begin{appendix}

\section{Installation of the relevant packages}\label{sec:installation}

The recommended installation of the relevant packages is done through
the \pkg{Pkg.jl}\footnote{\url{https://docs.julialang.org/en/v1/stdlib/Pkg/}} tool available within the \proglang{Julia} distribution standard
library. The common way to interact with the tool is within the REPL for
\proglang{Julia} 1.0 and above. The closing square bracket \code{"]"} starts
the \code{pkg} mode, in which the REPL stays until a return key is stroke.

\begin{verbatim}
julia> ] add StatsBase
(v1.0) pkg> add Distributions
(v1.0) pkg> add KernelDensity
\end{verbatim}

This installation and all code snippets are guaranteed to work on
\proglang{Julia} 1.0 and later with the semantic versioning commitment. Once
installed, a package can be removed with the command \code{rm PackageName}
or updated. All package modifications are guaranteed to respect version
constraints for their direct and transitive dependencies.
The packages \pkg{StatsBase.jl}, \pkg{Distributions.jl}, \pkg{KernelDensity.jl}
are all open-source under the MIT license.

\section[Julia type system]{\proglang{Julia} type system}\label{sec:apptype}

\subsection{Functions and methods}

In \proglang{Julia}, a function ``is an object that maps a tuple of argument
values to a return value''.\footnote{\proglang{Julia} documentation - Functions: \url{https://docs.julialang.org/en/v1/manual/functions}}
A special case is an empty tuple as input, as in \code{y = f()},
and a function that returns the \code{nothing} value.

A function definition creates a top-level identifier with the
function name. This can be passed around as any other value,
for example to other functions. The function \code{map} takes
a function and a collection \code{map(f, c)}, and applies the
function to each element of the collection to return the mapped
values.

A function might represent a conceptual computation but different
specific implementations might exist for this computation.
For instance, the addition of two numbers is a common concept, but
how it is implemented depends on the number type. The specific
implementation of addition for complex numbers is\footnote{\proglang{Julia} source code, \code{base/complex.jl}}
\begin{lstlisting}[language=Julia]
julia> +(z::Complex, w::Complex) = Complex(real(z) + real(w), imag(z) + imag(w))
\end{lstlisting}
This specific implementation is a \textbf{method} of the \code{+}
function. Users can define their own implementation of existing
functions, thus creating a new method for this function.
Different methods can be implemented by changing the tuple of arguments,
either with a different number of arguments or different
types.\footnote{\proglang{Julia} documentation - Methods: \url{https://docs.julialang.org/en/v1/manual/methods}}

\begin{example}
In the following example, the function \code{f} has two methods.
The function called depends on the number of arguments.
\begin{lstlisting}[language=Julia]
julia> function f(x)
    println(x)
end
julia> function f(x, y)
    println("x: $x & y: $y")
end
\end{lstlisting}
\end{example}

\begin{example}
In the following example, the function \code{g} has two methods.
The first one is the most specific method and will be called
for any type of the argument \code{x} that is a \code{Number}.
Otherwise, the second method, which is less specific, will be
called.
\begin{lstlisting}[language=Julia]
julia> g(x::T) where {T<:Number} = (3*x, x)
julia> g(x) = (x, 3)
\end{lstlisting}
Note that the order of definitions does not matter here, the least
specific could have been defined first, and then the number-specialized
implementation.
\end{example}

The method dispatched on by the \proglang{Julia} runtime is always
the most specific.
\begin{example}
If there is no unique most specific method, \proglang{Julia} will raise a \code{MethodError}
\begin{lstlisting}[language=Julia]
julia> f(x, b::Float64) = x
f (generic function with 1 method)
julia> f(x::Float64, b) = b
f (generic function with 2 methods)
julia> f(3.0, 2.0)
ERROR: MethodError: f(::Float64, ::Float64) is ambiguous. Candidates:
  f(x::Float64, b) in Main at REPL[2]:1
  f(x, b::Float64) in Main at REPL[1]:1
Possible fix, define
  f(::Float64, ::Float64)
\end{lstlisting}
\end{example}

\subsection{Types}

\proglang{Julia} enables users to define their own types
including abstract types, mutable and immutable composite
types or structures and primitive types (composed of bits).
Packages often define abstract types to regroup types
under one label and provide default implementation
for a group of types. For examples, lots of
methods require arguments which are identified as \code{Number},
upon which arithmetic operations can be applied,
without having to re-define methods for each of the
concrete number types. The most common type definition
for end-users is composite types with the keyword \code{struct}
as follows:
\begin{lstlisting}[language=Julia]
julia> struct S
    field1::TypeOfField1
    field2::TypeOfField2
    field3 # a field without specified type will take the type 'Any'
end
\end{lstlisting}
In some cases, a type is defined for the sole purpose of creating
a new method and does not require additional data in fields.
The following definition is thus valid:
\begin{lstlisting}[language=Julia]
julia> struct S end
\end{lstlisting}
An instance of \code{S} is called a \textbf{singleton}, there
is only one instance of such type.
\begin{example}
One use case is the specification of different algorithms for a
procedure. The input of the procedure is always the same,
so is the expected output, but different ways to compute
the result are available.
\begin{lstlisting}[language=Julia]
julia> struct Alg1 end
julia> struct Alg2 end
julia> mul(x::Unsigned, y::Unsigned, ::Alg1) = x * y
julia> mul(x::T, y::Unsigned, ::Alg2) where {T<:Unsigned} = T(sum(x for _ in 1:y))
\end{lstlisting}
\end{example}
Note that in the second example, the information of the
concrete type \code{T} of \code{x} is required to convert
the sum expression into a number of type \code{T}.

\section[Comparison of Distributions.jl, Python/SciPy and R]{Comparison of \pkg{Distributions.jl}, \proglang{Python}/\pkg{SciPy} and \proglang{R}}\label{sec:examples}

In this section, we develop several comparative examples
of how various computation tasks linked with probability distributions are
performed using \proglang{R}, \proglang{Python} with \pkg{SciPy/NumPy} and \proglang{Julia}.

\subsection{Sampling from various distributions}

The following programs draw 100 samples from a
Gamma distribution $\Gamma(k = 10, \theta = 2)$,
in \proglang{R}:
\begin{lstlisting}[language=Julia]
R> set.seed(42)
R> rgamma(100, shape = 10, scale = 2)
\end{lstlisting}
The following \proglang{Python} version uses the \pkg{NumPy} \& \pkg{SciPy} libraries.
\begin{lstlisting}[language=Julia]
>>> import numpy as np
>>> import scipy.stats as stats
>>> np.random.seed(42)
>>> stats.gamma.rvs(10, scale = 2, size = 100)
\end{lstlisting}
The following \proglang{Julia} version is written to stay close to
the previous ones, thus setting the global seed and not
passing along a new RNG object.
\begin{lstlisting}[language=Julia]
julia> import Random
julia> using Distributions
julia> Random.seed(42)
julia> g = Gamma(10, 2)
julia> rand(g, 100)
\end{lstlisting}

\subsection{Representing a distribution with various scale parameters}

The following code examples are in the same order as the previous
sub-section:
\begin{lstlisting}[language=Julia]
R> x <- seq(-3.0,3.0,0.01)
R> y <- dnorm(x, mean = 0.5, sd = 0.75)
R> plot(x, y)
\end{lstlisting}
The \proglang{Python} example uses \pkg{NumPy}, \pkg{SciPy} and \pkg{matplotlib}.
\begin{lstlisting}[language=Julia]
>>> import numpy as np
>>> from scipy import stats
>>> from matplotlib import pyplot as plt
>>> x = np.arange(-3.0,3.0,0.01)
>>> y = stats.norm.pdf(x, loc = 0.5, scale = 0.75)
>>> plt.plot(x,y)
\end{lstlisting}
The \proglang{Julia} version uses defines a Gaussian random variable
object \code{n}:
\begin{lstlisting}[language = Julia]
julia> using Distributions
julia> using Plots
julia> n = Normal(0.5, 0.75)
julia> x = -3.0:0.01:3.0
julia> y = pdf.(n, x)
julia> plot(x, y)
\end{lstlisting}
Note that unlike the \proglang{R} and \proglang{Python} examples,
x does not need to be represented as an array but as an iterable
not storing all intermediate values, saving time and memory.
The dot or broadcast operator in \code{pdf.(n, x)} applies
the function to all elements of \code{x}.

\section{Wine data analysis}\label{sec:wineanalysis}

In this section, we show the example of analyses run on the wine
dataset \cite{aeberhard1994comparative} obtained on the UCI
machine learning repository. The data can be fetched directly
over HTTP from within \proglang{Julia}.
\begin{lstlisting}[language = Julia]
julia> using Distributions, DelimitedFiles
julia> import Plots
julia> wine_data_url =
    "https://archive.ics.uci.edu/ml/machine-learning-databases/wine/wine.data"
julia> wine_data_file = download(wine_data_url)
julia> raw_wine_data = readdlm(wine_data_file, ',', Float64)
julia> wine_quant = raw_wine_data[:,2:end]
julia> wine_labels = Int.(raw_wine_data[:,1])
\end{lstlisting}

\subsection{Automatic multivariate fitting}\label{sec:winefitting}
We can then fit a multivariate distribution to some variables and observe
the result:
\begin{lstlisting}[language = Julia]
julia> alcohol = wine_quant[:,1]
julia> log_malic_acid = log.(wine_quant[:,2])
julia> obs = permutedims([alcohol log_malic_acid])
julia> res_normal = fit_mle(MvNormal, obs)
julia> cont_func(x1, x2) = pdf(res_normal, [x1,x2])
julia> p = Plots.contour(11.0:0.05:15.0, -0.5:0.05:2.5, cont_func)
julia> Plots.scatter!(p, alcohol, log_malic_acid, label="Data points")
julia> Plots.title!(
    p, "Wine scatter plot & Gaussian maximum likelihood estimation")
\end{lstlisting}
Note that \code{fit_mle} needs observations as columns, hence the use of \code{permutedims}.

\begin{figure}[t!]
    \centering
    \includegraphics[width = 0.6\textwidth]{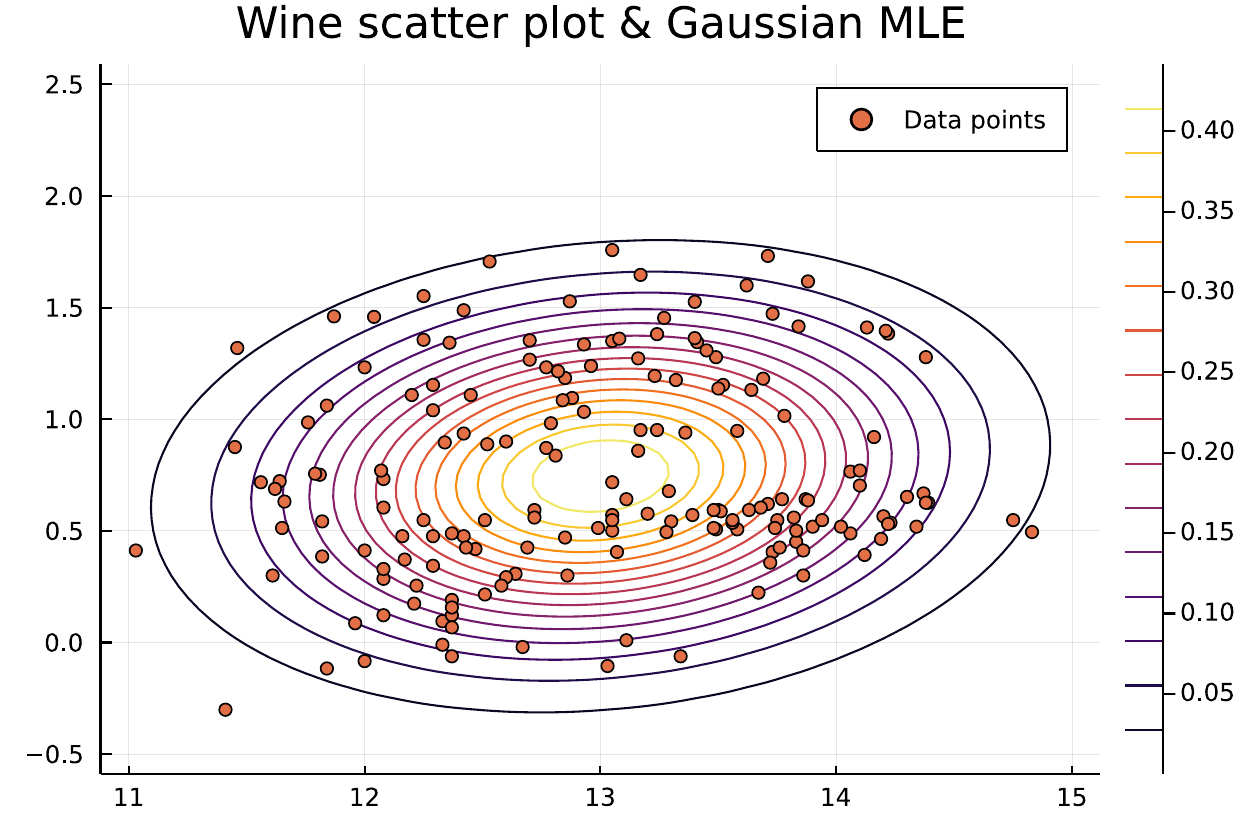}
    \caption{Multivariate Gaussian MLE over $(x_1, x_2)$ of the wine data}
    \label{fig:gaussian_contour_mle}
\end{figure}

\subsection{Non-parametric density estimation}\label{sec:winekde}

Assuming a multivariate Gaussian distribution might be considered a too strong
assumption, a kernel density estimator can be used on the same data:
\begin{lstlisting}[language = Julia]
julia> wine_kde = KernelDensity.kde((alcohol, log_malic_acid))
julia> cont_kde(x1, x2) = pdf(wine_kde, x1, x2)
julia> p = Plots.contour(11.0:0.05:15.0, -0.5:0.05:2.5, cont_kde)
julia> Plots.scatter!(p, alcohol, log_malic_acid, group=wine_labels)
julia> Plots.title!(p, "Wine scatter plot & Kernel Density Estimation")
\end{lstlisting}
\begin{figure}[t!]
\centering
\includegraphics[width = 0.6\textwidth]{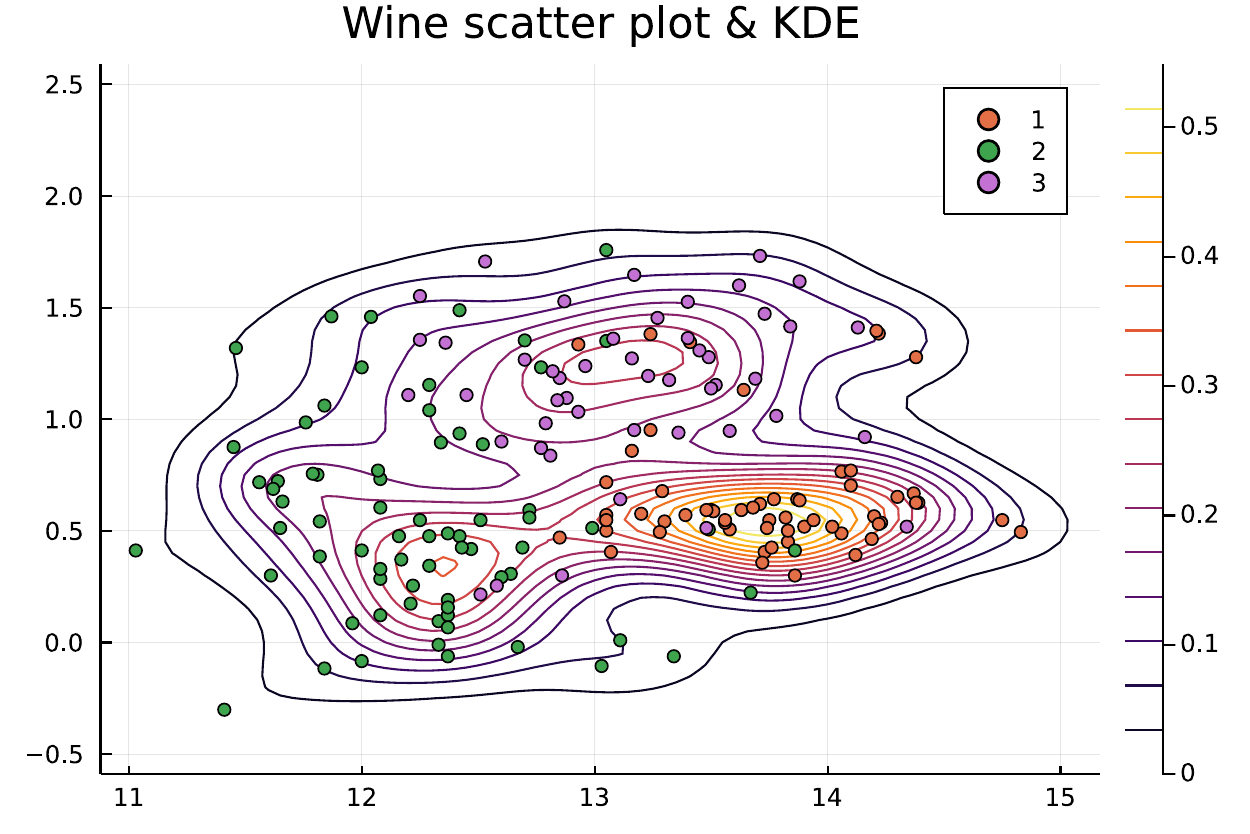}
\caption{KDE over $(x_1, x_2)$ of the wine data}
\label{fig:gaussian_contour_kde}
\end{figure}
The resulting kernel density estimation is shown Figure~\ref{fig:gaussian_contour_kde}. The level curves highlight the centers of the three classes of the dataset.

\subsection{Product distribution model}\label{sec:wineproduct}

Given that a logarithmic transform was used for $x_2$, a shifted log-normal
distribution can be fitted: $X_2 - 0.73 \sim  LogNormal(\mu,\sigma)$.
The maximum likelihood estimator could be computed as in section \ref{sec:winefitting}.
Instead, we will demonstrate the simplicity of building new constructs
and optimize over them. A \code{Product} distribution is implemented
in \pkg{Distributions.jl}, defining a Cartesian product of the two variables,
thus assuming their independence:
\begin{equation}
X = (X_1, X_2)
\end{equation}

\begin{figure}[t!]
\centering
\includegraphics[width = 0.8\textwidth]{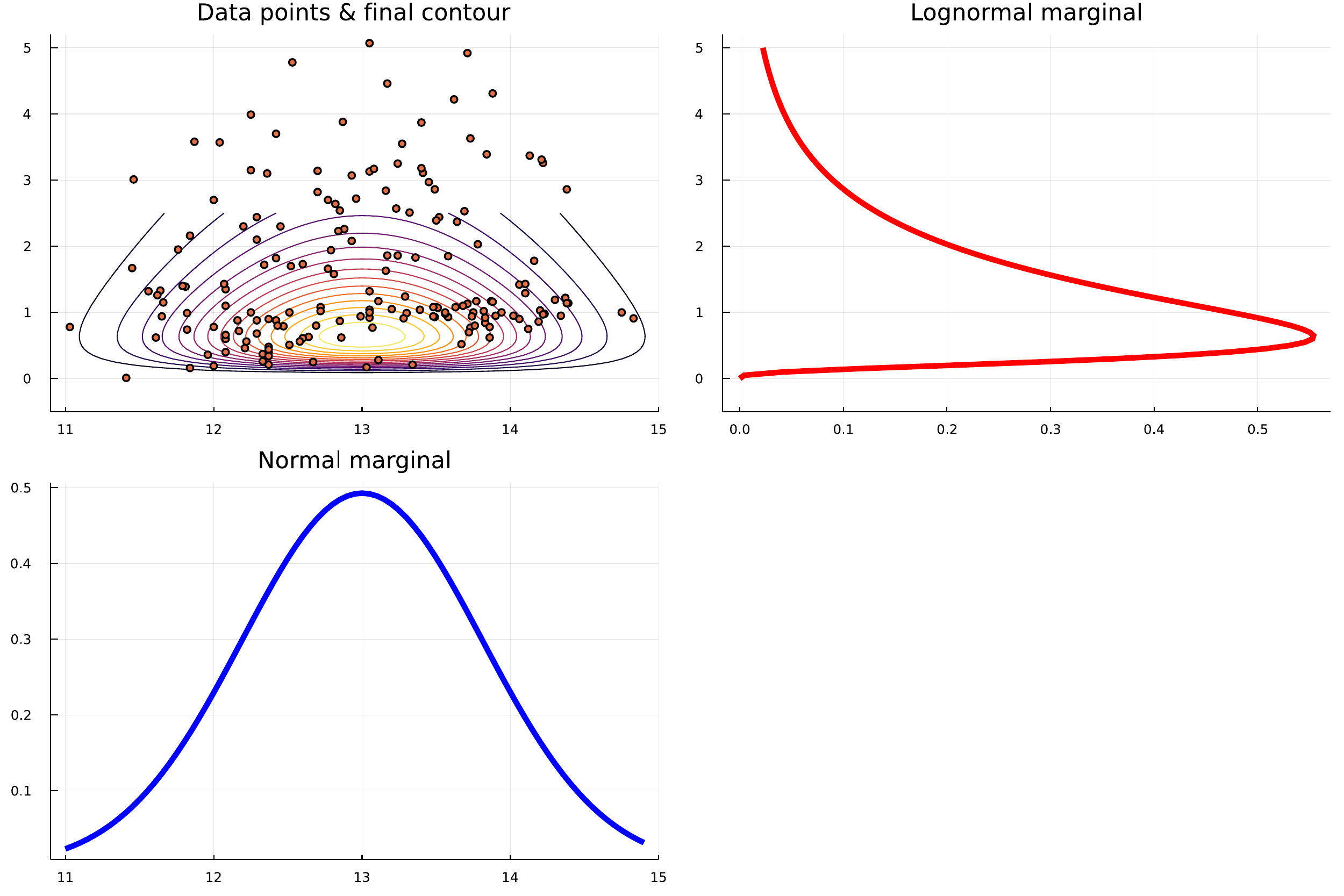}
\caption{Resulting inferred joint and marginal distribution for $(X_1, X_2)$}
\label{fig:wine_product_dist}
\end{figure}

\begin{figure}[t!]
\centering
\includegraphics[width = 0.6\textwidth]{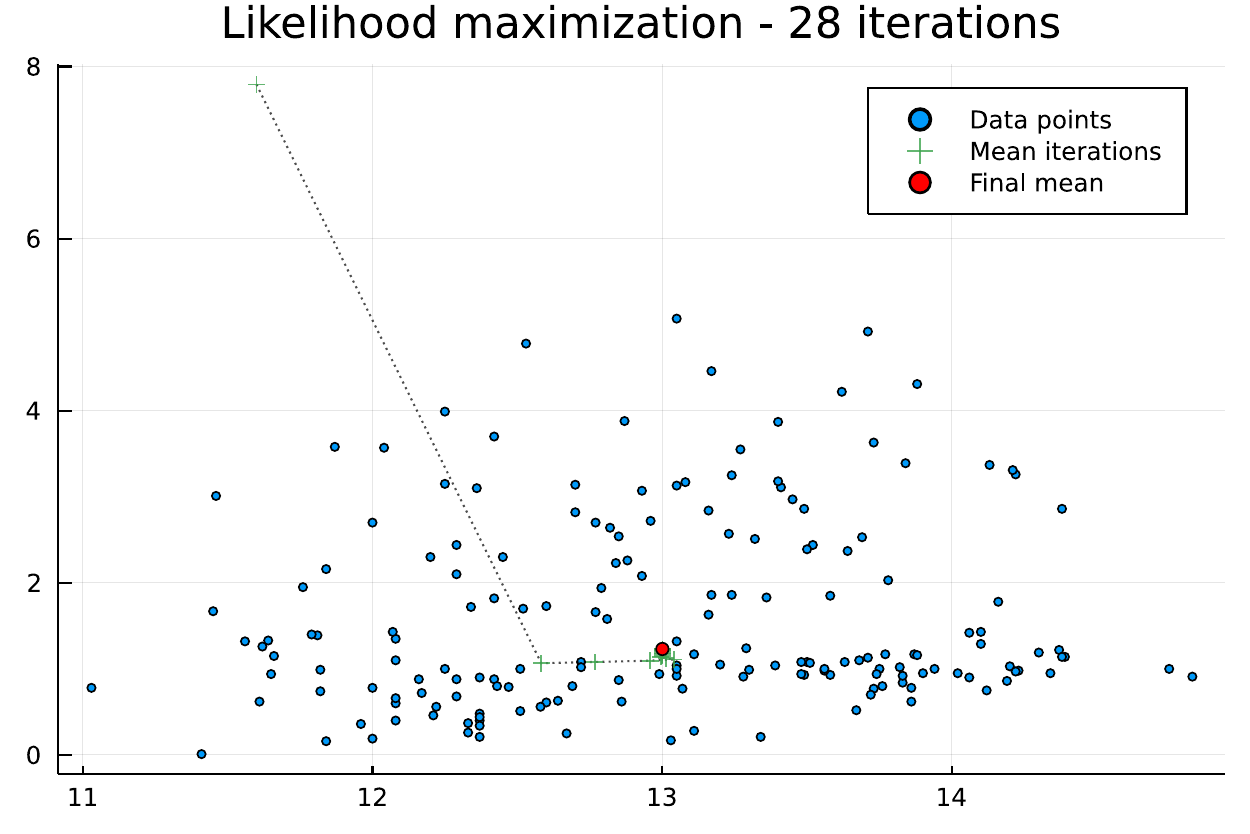}
\caption{Illustration of the likelihood maximization convergence}
\label{fig:wine_product_convergence}
\end{figure}

Assuming $X_1$ follows a normal distribution, given a vector of 4
parameters $\left[\mu_1,\sigma_1,\mu_2,\sigma_2\right]$, the distribution
can be constructed:
\begin{lstlisting}[language = Julia]
julia> function build_product_distribution(p)
    return product_distribution([
        Normal(p[1], p[2]),
        LogNormal(p[3], p[4]),
    ])
end
\end{lstlisting}
Computing the log-likelihood of a product distribution boils down
to the sum of the individual log-likelihood. The gradient could be
computed analytically but automatic differentiation will be leveraged
here using \pkg{ForwardDiff.jl} \citep{revels2016forward}:
\begin{lstlisting}[language = Julia]
julia> function loglike(p)
    d = build_product_distribution(p)
    return loglikelihood(d.v[1], wine_quant[:,1]) +
           loglikelihood(d.v[2], wine_quant[:,2] .- 0.73)
end
julia> ∇L(p) = ForwardDiff.gradient(loglike, p)
\end{lstlisting}
Once the gradient obtained, first-order optimization methods can be applied.
To keep everything self-contained, a gradient descent with decreasing
step will be applied:
\begin{align}
& x_{k+1} = x_{k} + \rho_k \nabla\mathcal{L}(x_k) \\
& \rho_{k+1} = \rho_{0} / (k + m)
\end{align}
With $(\rho_0, m)$ constants.
\begin{lstlisting}[language = Julia]
julia> p = [10.0 + 3.0 * rand(), rand()+1, 2.0 + 3.0*rand(), rand()+1]
julia> iter = 1
julia> maxiter = 5000
julia> while iter <= maxiter && sum(abs.(∇L(p))) >= 10^-6
    p = p + 0.05 * inv(iter+5) * ∇L(p)
    p[2] = p[2] < 0 ? -p[2] : p[2]
    p[4] = p[4] < 0 ? -p[4] : p[4]
    iter += 1
end
julia> d = build_product_distribution(p)
\end{lstlisting}
Without further code or method tuning, the optimization takes about
$590\mu s$ and few iterations to converge as shown Figure \ref{fig:wine_product_convergence}.
For more complex use cases, users would look at
more sophisticated techniques as developed in \pkg{JuliaSmoothOptimizers}
\citep{jso-2019} or \pkg{Optim.jl} \citep{mogensen2018optim}.
Figure \ref{fig:wine_product_dist} highlights the resulting marginal and
joint distributions.

\subsection{Implementation of an Expectation-Maximization algorithm}\label{sec:wineemalg}

In this section, we highlight how \proglang{Julia} dispatch mechanism
and \pkg{Distributions.jl} type hierarchy can help users define
algorithms in a generic fashion, leveraging specific structure but
allowing their extension. We know that the observations from
the wine data set are split into different classes $Z$.
We will consider only the two first variables, with the second
at a log-scale: $X = (x_1,log(x2))$.

The expectation step computes the probability for
each observation $i$ to belong to each label $k$:
\begin{lstlisting}[language = Julia]
julia> function expectation_step(X, dists, prior)
    n = size(X, 1)
    Z = Matrix{Float64}(undef, n, length(prior))
    for k in eachindex(prior)
        for i in 1:n
            Z[i,k] = prior[k] * pdf(dists[k], X[i,:]) /
                     sum(
                        prior[j] * pdf(dists[j], X[i,:])
                        for j in eachindex(prior)
                    )
        end
    end
    return Z
end
\end{lstlisting}
\code{dists} is a vector of distributions, note that
no assumption is needed on these distributions, other than
the standard interface. The only computation applied is indeed
the computation of the \code{pdf} for each of these.

The operation for which the specific structure of the distribution is
required is the maximization step. For many distributions, a closed-form
solution of the maximum-likelihood estimator is known, avoiding expensive
optimization schemes as developed in \ref{sec:wineproduct}. In the case
of the Normal distribution, the maximum likelihood estimator is given
by the empirical mean and covariance matrix. We define the function
\code{maximization_step} to take a distribution type, the data \code{X} and
current label estimates $Z$, computing both the prior probabilities of each
of the classes $\pi_k$ and the corresponding conditional distribution $D_k$,
it returns the pair of vectors $(D, \pi)$.
\begin{lstlisting}[language = Julia]
julia> function maximization_step(::Type{<:MvNormal}, X, Z)
    n = size(X, 1)
    Nk = size(Z, 2)
    µ = map(1:Nk) do k
        num = sum(Z[i,k] .* X[i,:] for i in 1:n)
        den = sum(Z[i,k] for i in 1:n)
        num / den
    end

    Σ = map(1:Nk) do k
        num = zeros(size(X, 2), size(X, 2))
        for i in 1:n
            r = X[i,:] .- µ[k]
            num .= num .+ Z[i,k] .* (r * r')
        end
        den = sum(Z[i,k] for i in 1:n)
        num ./ den
    end
    prior = [inv(n) * sum(Z[i,k] for i in 1:n)
            for k in 1:Nk
        ]
    dists = map(1:Nk) do k
        MvNormal(µ[k], Σ[k] + 10e-7I)
    end
    return (dists, prior)
end
\end{lstlisting}
The final block is a function alternatively computing $Z$ and $(D,\pi)$ until convergence:
\begin{lstlisting}[language = Julia]
julia> function expectation_maximization(
        D::Type{<:Distribution},
        X, Nk,
        maxiter = 500,
        loglike_diff = 10e-5)
    # initialize classes
    n = size(X,1)
    Z = zeros(n, Nk)
    for i in 1:n
        j0 = mod(i,Nk)+1
        j1 = j0 > 1 ? j0-1 : 2
        Z[i,j0] = 0.75
        Z[i,j1] = 0.25
    end
    (dists, prior) = maximization_step(D, X, Z)
    l = loglike_mixture(X, dists, prior)
    lprev = 0.0
    iter = 0
    # EM iterations
    while iter < maxiter && abs(lprev-l) > loglike_diff
        Z = expectation_step(X, dists, prior)
        (dists, prior) = maximization_step(D, X, Z)
        lprev = l
        l = loglike_mixture(X, dists, prior)
        iter += 1
    end
    return (dists, prior, Z, l, iter)
end
julia> function loglike_mixture(X, dists, prior)
    l = zero(eltype(X))
    n = size(X,1)
    for i in 1:n
        l += log(
            sum(prior[k] * pdf(dists[k], X[i,:]) for k in eachindex(prior))
        )
    end
    return l
end
\end{lstlisting}
Starting from an alternated affectation of observations to labels, it calls the two methods
defined above for the \textit{E} and \textit{M} steps.

\begin{figure}[t!]
\begin{tabular}{cc}
  \includegraphics[width=65mm]{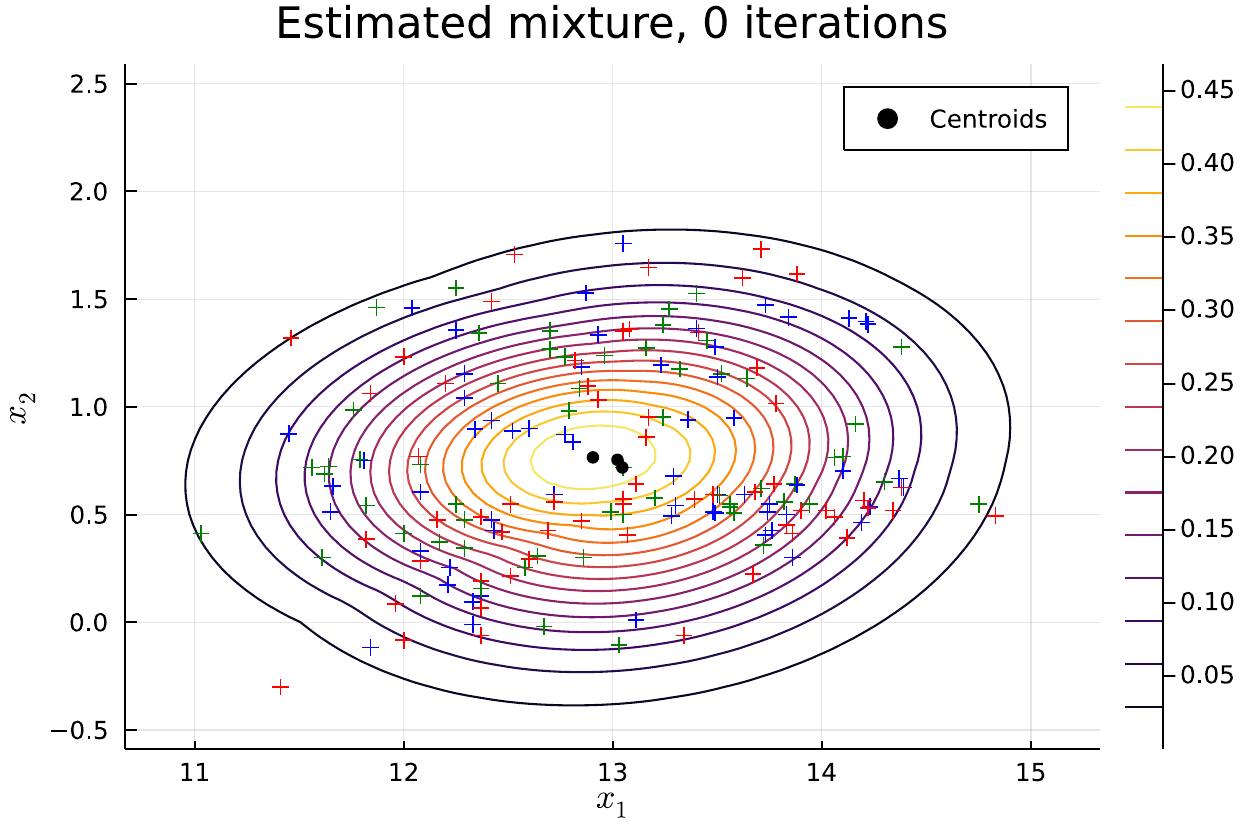} &   \includegraphics[width=65mm]{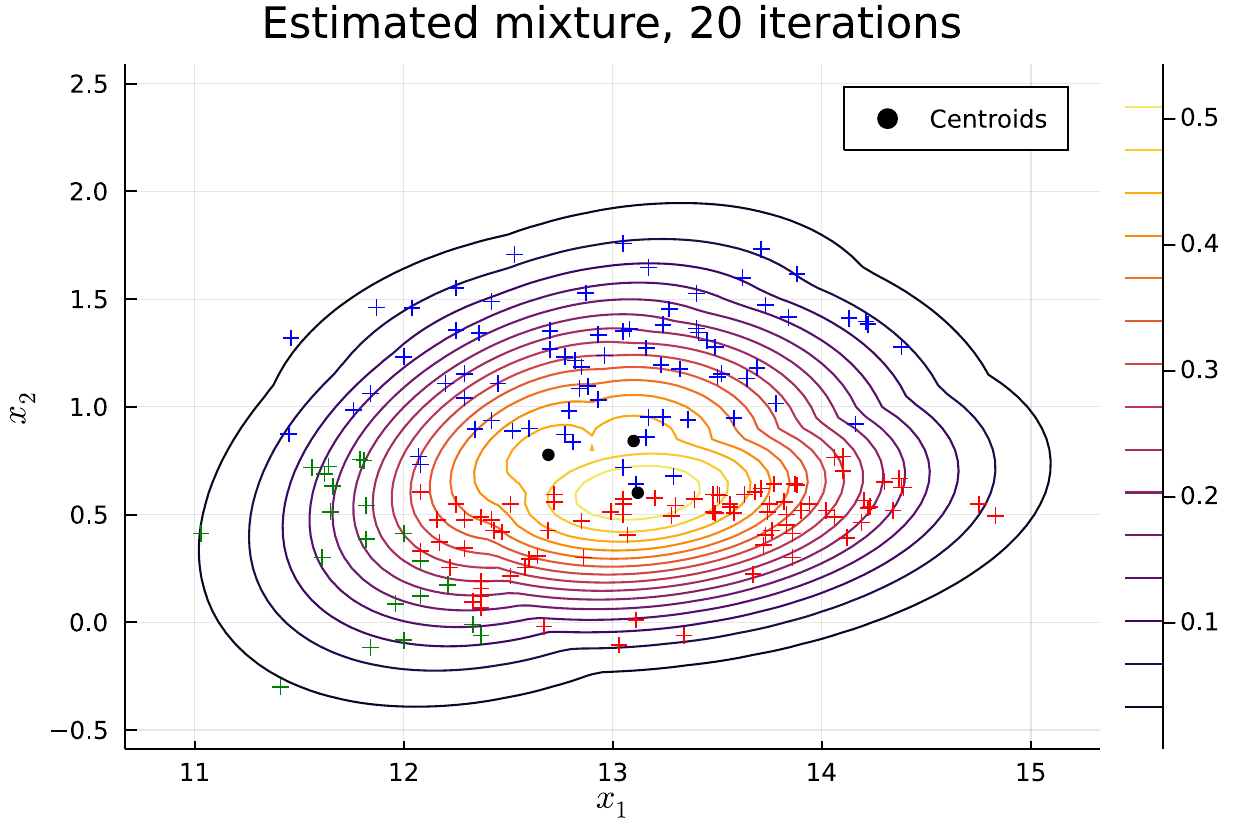} \\
$n = 0$ & $n=20$ \\[6pt]
 \includegraphics[width=65mm]{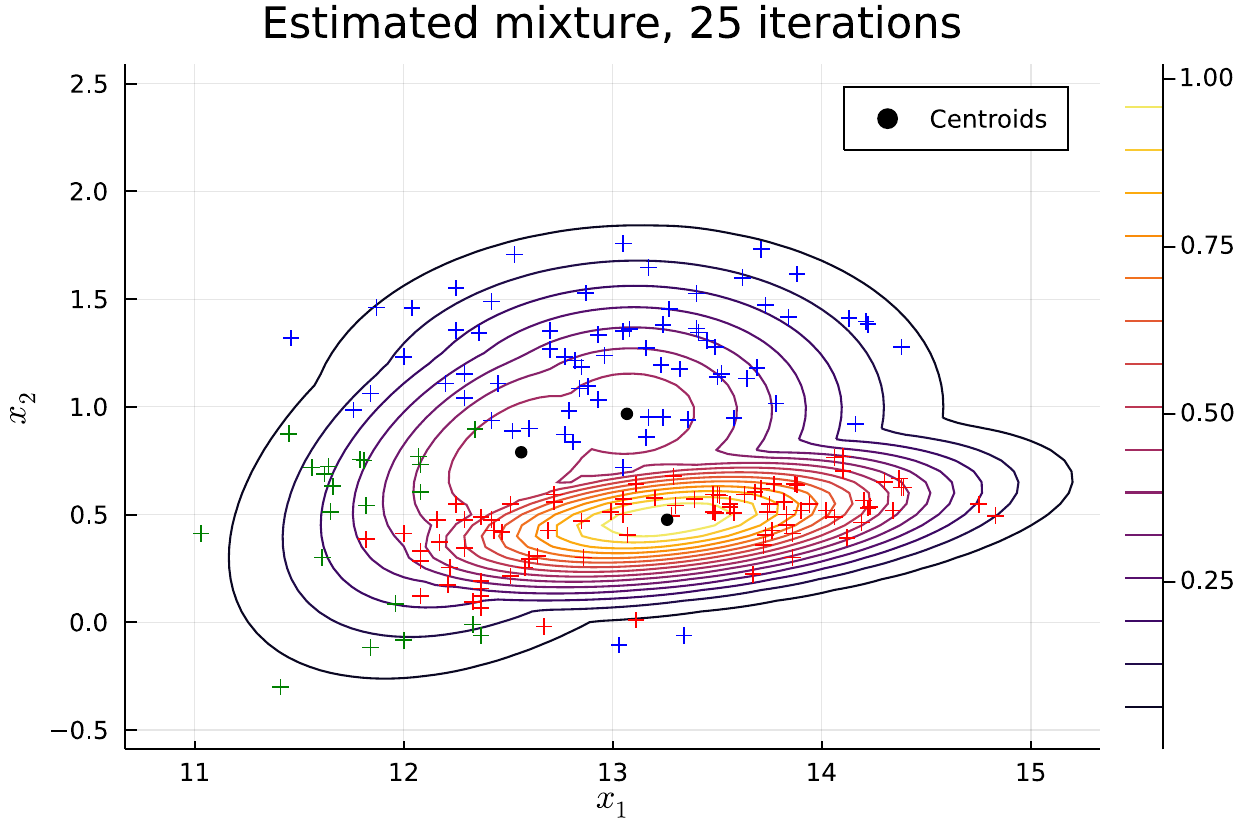} &   \includegraphics[width=65mm]{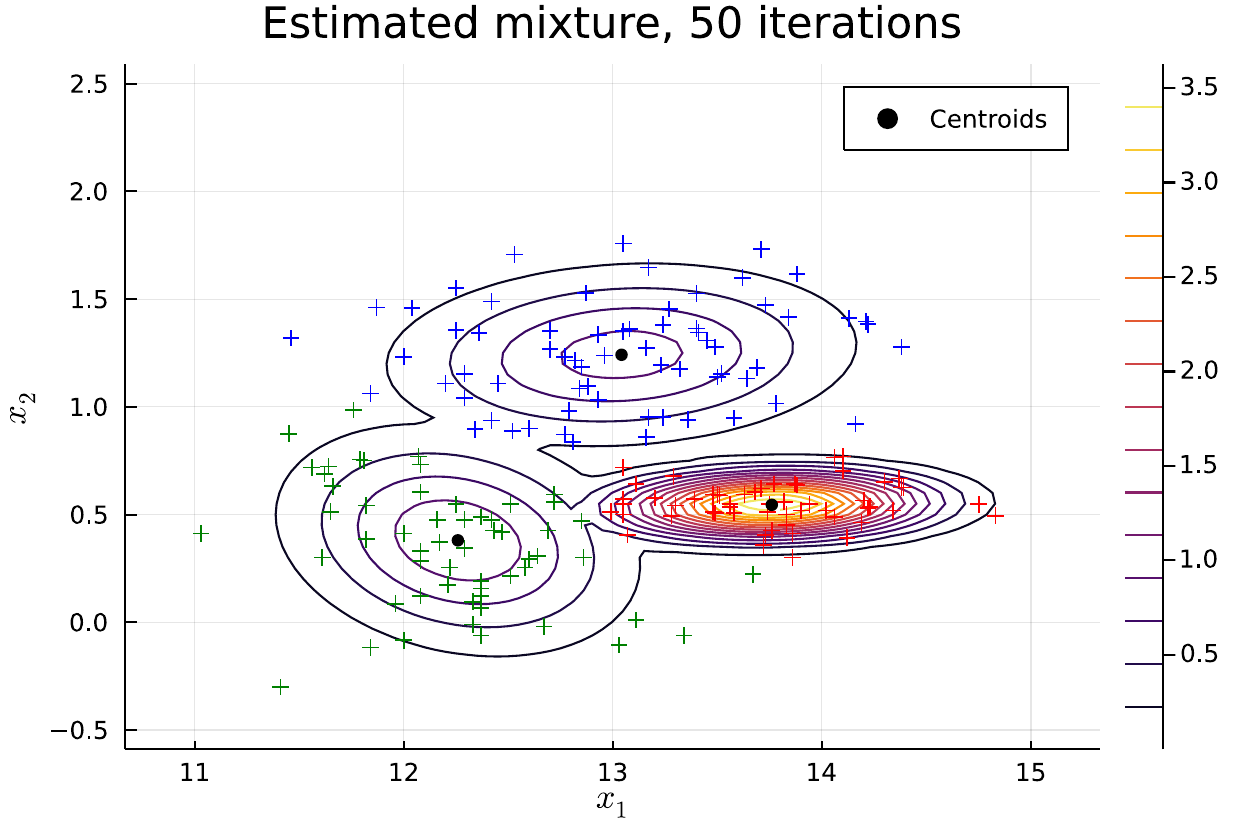}\\
 $n = 25$ & $n=50$ \\[6pt]
 \end{tabular}
\caption{Convergence of the EM algorithm}
\end{figure}

The strength of this implementation is that extending it to more functions only
requires to implement \code{maximization_step(D, X, Z)} for the new distribution.
The rest of the structure will then work as expected.

\section{Benchmarks on mixture PDF evaluations}

In this section, we evaluate the computation time for the PDF of mixtures of
distributions, on mixture of Gaussian univariate distributions, and then
on a mixture of heterogeneous distributions including normal and log-normal
distributions.
\begin{lstlisting}[language = Julia]
julia> using BenchmarkTools: @btime
julia> distributions = [
    Normal(-1.0, 0.3),
    Normal(0.0, 0.5),
    Normal(3.0, 1.0),
]
julia> priors = [0.25, 0.25, 0.5]
julia> gmm_normal = MixtureModel(distributions, priors)
julia> x = rand()
julia> @btime pdf($gmm_normal, $x)
julia> large_normals = [Normal(rand(), rand()) for _ in 1:1000]
julia> large_probs = [rand() for _ in 1:1000]
julia> large_probs .= large_probs ./ sum(large_probs)
julia> gmm_large = MixtureModel(large_normals, large_probs)

@btime pdf($gmm_large, $x)

julia> large_h = append!(
    ContinuousUnivariateDistribution[Normal(rand(), rand()) for _ in 1:1000],
    ContinuousUnivariateDistribution[LogNormal(rand(), rand()) for _ in 1:1000],
)
julia> large_h_probs = [rand() for _ in 1:2000]
julia> large_h_probs .= large_h_probs ./ sum(large_h_probs)
julia> gmm_h = MixtureModel(large_h, large_h_probs)
julia> @btime pdf($gmm_h, $x)
\end{lstlisting}
The PDF computation on the small, large and heterogeneous mixture take on average
$332ns$, $105\mu s$ and $259\mu s$ respectively.
We also compare the computation time with the manual computation of the mixture
evaluated as such:
\begin{lstlisting}[language = Julia]
julia> function manual_computation(mixture, x)
    v = zero(eltype(mixture))
    p = probs(mixture)
    d = components(mixture)
    for i in 1:ncomponents(mixture)
        if p[i] > 0
            v += p[i] * pdf(d[i], x)
        end
    end
    return v
end
\end{lstlisting}
The PDF computation currently in the library is faster than the manual one,
the sum over weighted PDF of the components being performed without branching.
The ratio of manual time over the implementation of \pkg{Distributions.jl}
being on average 1.24, 5.88 and 1.28 respectively.
The greatest acceleration is observed for large mixtures with homogeneous
component types. The slow-down for heterogeneous component types is due to the
compiler not able to infer the type of each element of the mixture, thus replacing
static with dynamic dispatch.
The \pkg{Distributions.jl} package also contains a \code{/perf} folder containing
experiments on performance to track possible regressions and compare different
implementations.

\end{appendix}

\end{document}